\documentclass[reprint,amsmath,amssymb,aps,superscriptaddress,longbibliography]{revtex4-2}
\usepackage{graphicx}
\usepackage{subfigure}
\usepackage{comment}
\usepackage{dcolumn}
\usepackage{bm}
\usepackage{amsmath,amssymb}
\usepackage{amsthm}
\usepackage{mathrsfs}
\usepackage{indentfirst}
\usepackage{float}
\usepackage{braket}
\usepackage{physics}
\usepackage[utf8]{inputenc}
\usepackage{bm}
\usepackage{graphicx}
\usepackage{tikz}
\usepackage{physics}
\usetikzlibrary{arrows, shapes.gates.logic.US, calc}
\usetikzlibrary{angles,quotes}
\tikzstyle{branch}=[fill, shape=circle, minimum size=3pt, inner sep=0pt]
\usepackage{blochsphere}
\usepackage{mathtools}
\usepackage{color,graphicx} 
\newcommand\D{\!\operatorname{d}\!}
\usepackage{algorithm}
\usepackage[algo2e]{algorithm2e}
\usepackage{bbold}

\usetikzlibrary{quantikz}

\usepackage{hyperref}
\hypersetup{
 colorlinks=true,
 citecolor=red,
 linkcolor=blue,
 urlcolor=blue,
 pdfpagemode=UseNone,
 pdfstartview=FitH}
\usepackage{subfiles}

\newcommand{\orcidicon}[1]{\href{https://orcid.org/#1}{\includegraphics[height=\fontcharht\font`\B]{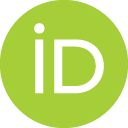}}}

\footnotetext{These authors contributed equally}

\begin{document}
\preprint{APS/123-QED}
\title{Efficient quantum algorithm to simulate open systems \\ through a single environmental qubit}

\author{Giovanni Di Bartolomeo\,\orcidicon{0000-0002-1792-7043}$^\diamond$}
\email{giovanni.dibartolomeo@phd.units.it}
\affiliation{Department of Physics, University of Trieste, Strada Costiera 11, 34151 Trieste, Italy}
\affiliation{Istituto Nazionale di Fisica Nucleare, Trieste Section, Via Valerio 2, 34127 Trieste, Italy}

\author{Michele Vischi\,\orcidicon{0000-0002-5724-7421}$^\diamond$}

\email{michele.vischi@phd.units.it}
\affiliation{Department of Physics, University of Trieste, Strada Costiera 11, 34151 Trieste, Italy}
\affiliation{Istituto Nazionale di Fisica Nucleare, Trieste Section, Via Valerio 2, 34127 Trieste, Italy}

\author{Tommaso~Feri\,\orcidicon{0009-0005-0320-2964}$^\diamond$}
\email{tommaso.feri@phd.units.it}
\affiliation{Department of Physics, University of Trieste, Strada Costiera 11, 34151 Trieste, Italy}
\affiliation{Istituto Nazionale di Fisica Nucleare, Trieste Section, Via Valerio 2, 34127 Trieste, Italy}

\author{Angelo Bassi\,\orcidicon{0000-0001-7500-387X}}
\affiliation{Department of Physics, University of Trieste, Strada Costiera 11, 34151 Trieste, Italy}
\affiliation{Istituto Nazionale di Fisica Nucleare, Trieste Section, Via Valerio 2, 34127 Trieste, Italy}

\author{Sandro Donadi\,\orcidicon{0000-0001-6290-5065}}
\affiliation{Centre for Quantum Materials and Technologies, School of Mathematics and Physics, Queen’s University, Belfast BT7 1NN, United Kingdom}
\begin{abstract}
We present an efficient algorithm for simulating open quantum systems dynamics described by the Lindblad master equation on quantum computers, addressing key challenges in the field. In contrast to existing approaches, our method achieves two significant advancements. First, we employ a repetition of unitary gates on a set of $n$ system qubits and, remarkably, only a single ancillary bath qubit representing the environment. It follows that, for the typical case of $m$-locality of the Lindblad operators, we reach an exponential improvement of the number of ancilla in terms of $m$ and up to a polynomial improvement in ancilla overhead for large $n$ with respect to other approaches. Although stochasticity is introduced, requiring multiple circuit realizations, the sampling overhead is independent of the system size. Secondly, we show that, under fixed accuracy conditions, our algorithm enables a reduction in the number of trotter steps compared to other approaches, substantially decreasing circuit depth. These advancements hold particular significance for near-term quantum computers, where minimizing both width and depth is critical due to inherent noise in their dynamics.
\end{abstract}

\maketitle

\section{Introduction}
Quantum systems are typically presented as isolated from the surrounding environment: this is often a good approximation in simple cases, such as for single photons or single atoms, but fails in more complex situations. The theory of open quantum systems, which takes into account the effects of the environment, has a long and celebrated history, with key results such as the Linblad-Gorini-Kossakowski-Sudarshan (LGKS) master equation \cite{gorini1976completely, lindblad1976generators}. In this context, as standard in physics, simple problems have been well understood theoretically, while complex ones require numerical simulations.
 
Simulating open quantum systems allows to model and understand real-world quantum phenomena and is crucial in a variety of fields: in quantum chemistry, understanding chemical reactions at a quantum level facilitates the design of new materials and the development of novel molecules \cite{mohseni2008environment,caruso2009highly,huelga2013vibrations,RevModPhys.92.015003}; in the field of condensed matter physics, such simulations are employed to explore the dynamics of quantum materials \cite{PhysRevE.90.042142,PhysRevA.91.022121, PhysRevA.86.012116,PhysRevA.88.042115,PhysRevLett.112.094102,PhysRevLett.113.154101}; in quantum optical systems, they help in understanding the behavior of quantum systems made of light, such as quantum lasers and quantum teleportation systems \cite{bruneau2014mixing,ciccarello2017collision,cilluffo2020collisional,carollo2020mechanism}; in quantum thermodynamics, they allow to explore energy exchanges, work extraction, and other thermodynamic processes in the quantum regime \cite{PhysRevLett.115.120403,pezzutto2016implications,PhysRevLett.102.207207,PhysRevE.96.052114,Lostaglio2018elementarythermal,PhysRevA.98.032119,PhysRevE.99.042103,arisoy2019thermalization,PhysRevE.99.042145,Korzekwa_2021,ehrich2020micro,guarnieri2020non,PhysRevA.78.012308,hofer2017markovian,gonzalez2017testing}. Finally, simulating open systems is relevant to the development of quantum technologies, including high-precision quantum sensors \cite{ciccarello2017collision,PhysRevLett.123.180602}, quantum batteries \cite{PhysRevLett.122.210601}, and quantum engines \cite{pezzutto2019out, PhysRevE.101.012109, PhysRevResearch.2.033315}. 
 
Quantum computers are a privileged tool for simulating open quantum systems, an idea which has gained increasing interest in recent years, showing several challenges that must be addressed. Endeavors in this direction primarily fall into two categories. The first one is based on the concept that, since the quantum devices themselves are open systems subject to noise, one can harness the inherent noise of the computer to simulate the desired open system. Recent promising results have emerged along this line, holding the potential for fruitful outcomes \cite{guimaraes2023noise,leppakangas2022quantum, sun2021efficient}. However, this approach is limited in that it can only simulate open systems described by the same master equation that characterizes the device and, moreover, building an adequate noise model of the device itself poses a significant challenge \cite{georgopoulos2021modeling,di2023noisy,vischi2023simulating,PhysRevApplied.20.034065,martina2022learning}. The second line of research is based on the ambition that also for open systems, efficient quantum algorithms can be devised, without relying on the natural noise of the computer \cite{kliesch2011dissipative,sweke2015universal,2017,de2021quantum,cattaneo2021collision, cattaneo2023quantum, cleve2016efficient,kamakari2022digital,pocrnic2024quantum}. The present work falls within the latter framework.
 
In this letter we propose an efficient quantum algorithm that simulates the Lindblad dynamics of the system density matrix $\hat{\rho}_{S}(t)$ \cite{gorini1976completely,lindblad1976generators}:
\begin{equation}
\label{Lindblad}
\begin{aligned}
\frac{\D}{\D t}\hat{\rho}_{S}(t)&=-\frac{i}{\hbar}[\hat{H}_{\text{S}},\hat{\rho}_{S}(t)]\\
&+\sum_k\gamma_k\Bigl(\hat{L}_k\hat{\rho}_{S}(t)\hat{L}_k^{\dagger}-\frac{1}{2}\{\hat{L}_k^{\dagger}\hat{L}_k,\hat{\rho}_{S}(t)\}\Bigr),
\end{aligned}
\end{equation}
where $\hat{H}_{\text{S}}$ is the system Hamiltonian, $\hat{L}_k$ are the Lindblad operators that describe the effects of the environment on the system and $\gamma_k$ are the coupling constants between the system and the environment. We reach two main significant advancements in the field.
 
First, we show that it is always possible to drive the open system dynamics via a repetition of unitary gates applied to a set of $n$ system qubits and crucially only a single ancillary bath qubit, representing the environment. As such, the ancilla overhead is always equal to 1, meaning that it does not scale with the number $n$ of qubits neither, in the case of $m$-locality of the Lindblad operators, with $m$. For a comparison, among state of the art implementations, some methods  \cite{sweke2015universal,de2021quantum,cattaneo2021collision, cattaneo2023quantum, cleve2016efficient,kamakari2022digital} use a number of ancillary qubits that scales polynomially with large $n$; our method provides a polynomial improvement, for large $n$, with respect to them. For other methods \cite{kliesch2011dissipative} the number of ancillary qubits scales exponentially with $m$; in this case we provide an exponential improvement.
The price to pay is stochasticity, i.e. multiple realizations of the quantum circuit, which however is preferable because the sampling overhead does not scale with the number of qubits.
 
Second, given a fixed accuracy, for small system-environment coupling constants, the approximations that we use allows to implement a number of trotter steps which is smaller than those required in other approaches \cite{cattaneo2021collision, cattaneo2023quantum,pocrnic2024quantum}, implying a reduction of the circuit depth.
 
These two advancements are important in general, and even more for near-term quantum computers where, due to their noisy dynamics, a drastic reduction of both width and depth of the quantum circuit is crucial.

\begin{figure}[t!]
\centering
\includegraphics[width=0.45\textwidth]{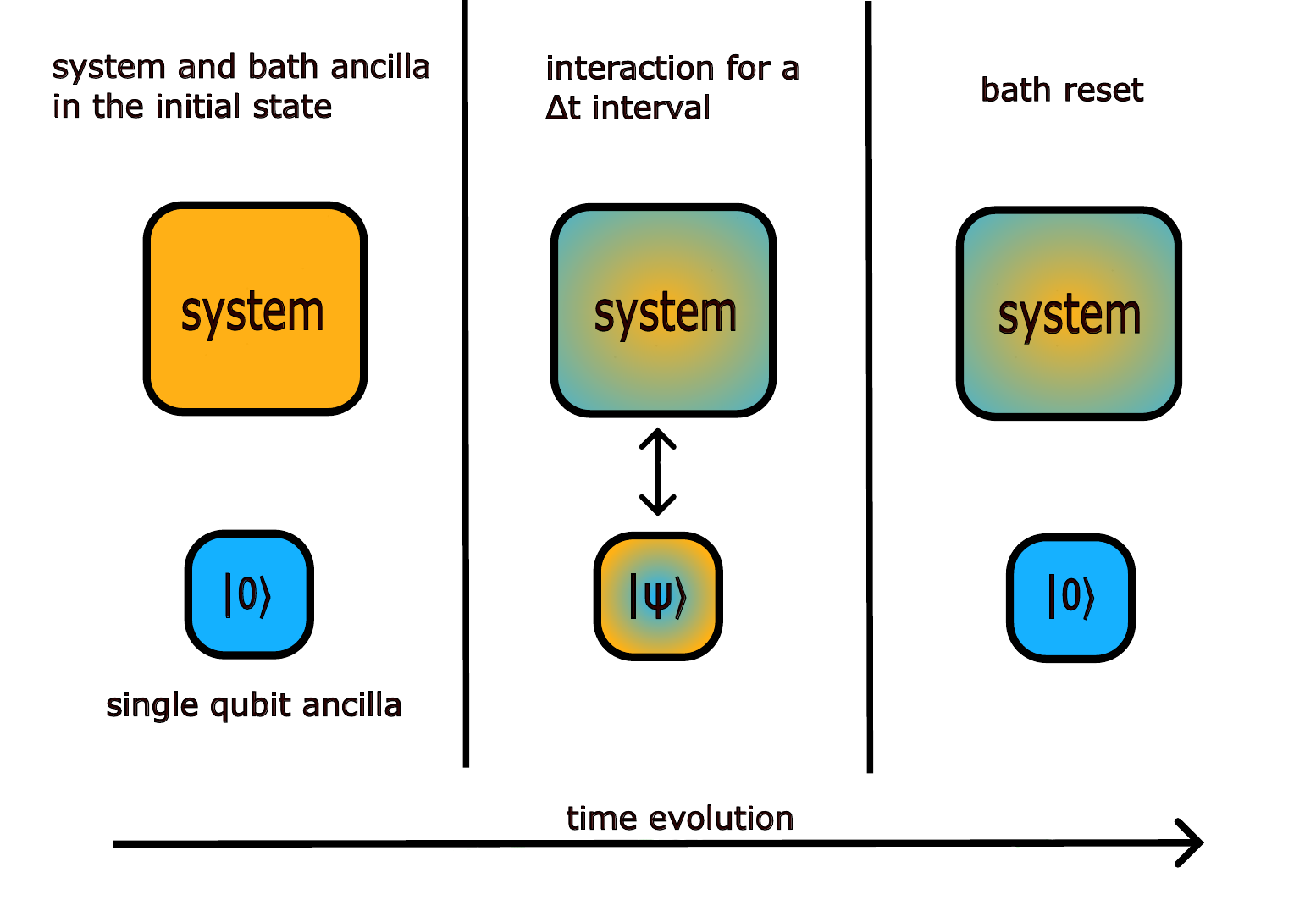}
\caption{Schematic picture of the three main steps of the protocol: preparation of the system and bath qubit in the initial state; system-bath interaction for a time $\Delta t$; reset of the bath qubit. In order to obtain the total evolution at a given time $T$, the last two steps have to be repeated $N_{steps} = T / \Delta t$ times.}
\label{disegno}
\end{figure}
\section{The quantum noise formalism}
Our approach is based on the quantum noise formalism \cite{gardiner2004quantum,wiseman2009quantum, zoller1997quantum}, that we introduce with more details in the appendix \ref{appendix_A} below. In this framework, one considers a system in contact with an environment made of thermal baths, each with annihilation operators $\hat{b}_k(\omega)$.
The evolution of the state vector $\ket{\Psi(t)}$ of the system plus environment  is described by the following quantum stochastic differential equation (QSDE) 
\begin{equation}
 \label{QSDE}
\begin{aligned}
   \D\ket{\Psi(t)}= &\Bigl[ -\frac{i}{\hbar}\hat{H}_{\text{S}}\D t + \sum_k\sqrt{\gamma_k}\left(\hat{L}_k\,\D \hat{B}_k^{\dagger}(t) - \hat{L}_k^{\dagger} \,\D \hat{B}_k(t) \right)\\ 
   &-\sum_k\frac{\gamma_k}{2} \hat{L}_k^{\dagger} \hat{L}_k dt \Bigr]\ket{\Psi(t)},
\end{aligned} 
\end{equation}
where $\D \hat{B}_k(t) = \hat{b}_{\text{in},k}(t)\D t$ with $\hat{b}_{\text{in},k}(t) = \frac{1}{\sqrt{2 \pi}} \int \D \omega e^{-i\omega(t-t_0)}\hat{b}_k(t_0,\omega)$.
The operators $\hat{B}_k(t)$ can be interpreted as the quantum generalization of a Wiener process \cite{gardiner1985handbook, jacobs2010stochastic} and their variances satisfy 
\begin{equation}\label{mean_quantum_wiener}
   \mathbb{E}_{\text{Q}}\Bigl[\D \hat{B}_k(t)\D\hat{B}_j^{\dagger}(t)\Bigr] = \delta_{kj}\D t 
\end{equation}
where $\mathbb{E}_{\tiny\text{Q}}[\,\cdot\,] \equiv \Tr_{\text{E}}(\hat{\rho}_{\text{in}}\,\cdot\,)$ with $\hat{\rho}_{\text{in}} = \ket{0_\omega}_{\text{E}}\bra{0_\omega}_{\text{E}}$ and $\ket{0_\omega}_{\text{E}}$ is the ground state of the thermal baths; $\Tr_{\text{E}}(\cdot)$ is the trace over the environment. The mean values of the other combinations of two thermal baths operators are zero.

 From Eq.~\eqref{QSDE} one can show (see Appendix ~\ref{standard_deriv}) that the master equation for the reduced density matrix  of the  system  $\hat{\rho}_{S}(t) = \Tr_{\text{E}}(\ket{\Psi(t)}\bra{\Psi(t)})$  is the Lindblad Eq.~\eqref{Lindblad}. 

\section{Approximate expression of the unitary evolution}\label{sec:approx_solution}
The unravelling in Eq. \eqref{QSDE} acting on the full system-environment space is linear meaning that knowing the state at time $t$, the state at time $t+\Delta t$ is given by
\begin{equation}
\label{state_evolution}
\ket{\Psi(t+\Delta t)} = \hat{N}(\Delta t)\ket{\Psi(t)},
\end{equation}
where $\hat{N}(\Delta t)$ is a matrix which is also unitary, thus implementable on a quantum computer. 
As we show in Appendix \ref{perturbative_solution_appendix}, the unitary evolution $\hat{N}(\Delta t)$ in Eq.\eqref{state_evolution} can be obtained by resorting to perturbative techniques \cite{gardiner1985handbook}, as Eq. \eqref{QSDE} in general is not solvable in a closed form. This leads to the following approximate expression
\begin{equation}
\begin{aligned}
\label{Q_noisy_gate}
\hat{N}(\Delta t) &= \hat{U}(\Delta t) \text{T} \big[ e^{\sum_k\sqrt{\gamma_k}\hat{S}_k(\Delta t)}\big]\\
&\simeq  \hat{U}(\Delta t) \prod_k e^{\sqrt{\gamma_k}\hat{S}_k(\Delta t)},
\end{aligned}
\end{equation}
where $\hat{U}(\Delta t)$ is the closed system evolution operator, $\text{T}[\cdot]$ is the time ordering operator and the stochastic terms $\hat{S}_k(t)$ are defined as
\begin{align}
    \hat{S}_k(\Delta t) = \int_t^{t+\Delta t} \Bigl(\hat{L}_k(s) d\hat{B}_k^{\dagger}(s) - \hat{L}_k^{\dagger}(s) d\hat{B}_k(s)\Bigr),
    \label{S_k}
\end{align}
and $\hat{L}_k(s) = \hat{U}^{\dagger}(s-t)\hat{L}_k\hat{U}(s-t)$ are the Lindblad operators in interaction picture.  

Under this approximation by computing $\hat{\rho}_{S}(t+\Delta t) = \Tr_{E}\big(\hat{N}(\Delta t)\ket{\Psi(t)}\bra{\Psi(t)}\hat{N}^{\dagger}(\Delta t)\big) $ one recovers the following final density matrix of the system (see Appendix \ref{proof_eq_7})
\begin{equation}
\begin{aligned}
\label{rho_ng_approx}
&\hat{\rho}_{S}(t+\Delta t)\simeq \hat{U}(\Delta t)\biggl(\hat{\rho}_{S}(t)+\int_{t}^{t+\Delta t} \D s\mathcal{D}(s)\hat{\rho}_{S}(t)\biggr)\hat{U}^{\dagger}(\Delta t)
\end{aligned}
\end{equation}
where 
$\mathcal{D}(s)\hat{\rho}_{S}(t)=\hat{U}^{\dagger}(s-t) \mathcal{D}\hat{\rho}_{S}(t)\hat{U}(s-t)=\sum_k\!\gamma_k\bigl(\hat{L}_k(s)\hat{\rho}_{S}(t)\hat{L}_k^{\dagger}(s)-\frac{1}{2}\bigl\{\hat{L}_k^{\dagger}(s)\hat{L}_k(s),\hat{\rho}_{S}(t)\bigr\}\bigr)$.
%
The approximate expression in Eq. \eqref{rho_ng_approx} leads to an approximation error due to the perturbative expansion
\begin{equation}
\label{e_p}
\begin{split}
\varepsilon_p &= \biggl\lVert\hat{U}(\Delta t)\text{T}\bigl[e^{\int_t^{t+\Delta t}\D s\mathcal{D}(s)}\bigr]\hat{U}^{\dagger}(\Delta t)\\
&-\hat{U}(\Delta t)\Bigl(\mathbb{1} +\int_t^{t+\Delta t}\D s\mathcal{D}(s)\Bigr)\hat{U}^{\dagger}(\Delta t)\biggr\rVert_{1\rightarrow 1}
\end{split}
\end{equation}
where we use the $1\rightarrow 1$
superoperator norm \cite{kliesch2011dissipative,sweke2015universal}. The term in the first line of Eq. \eqref{e_p} is the formal solution of the
Lindblad equation in interaction picture, where $\text{T}[\,\cdot\,]$ is the time ordering operator. $\varepsilon_p$ is quantified in Appendix \ref{approx_error_appendix} by assuming a $m$-locality condition, namely by considering the decomposition of $\mathcal{D}(s)$ into $\mathcal{D}(s) = \sum_{j}^{K}\mathcal{D}_j(s)$ where each $\mathcal{D}_j(s)$ acts non trivially on a subset of $m < n$ qubits. Then each $\mathcal{D}_j(s)$ has a maximum of $2^{2m}-1$ Lindblad operators. For simplicity we assume that all parameters $\gamma_k$ have the same order of magnitude $\gamma_k = \gamma$. Thus, the approximation error is bounded by
\begin{equation}
\label{bound_e_p}
\varepsilon_p \leq 2 e \Bigl(K (2^{2m}-1)\max_{k,j}\norm{\hat{L}_{k,j}}^{2}_{\infty}\gamma\Delta t\Bigr)^2, 
\end{equation}
where $e$ is the Euler number and $K(2^{2m}-1)$ is the total number of Lindblad operators which scales polynomially with the number of the system qubits $n$: $K \sim \mathcal{O}(n^m)$ (see Appendix \ref{approx_error_appendix}).

\section{Finite representation of the bath operators}\label{sec:finite_repr}
If we wish to simulate the evolution given by Eq. \eqref{state_evolution} on quantum computers, a problem arises in terms of how to represent the terms $\D \hat{B}_{k}(t)$ and $\D\hat{B}_{k}^{\dagger}(t)$,
which are operators acting on an infinite dimensional Hilbert space. To overcome this difficulty we proceed as follows: we
assume that at every given time $\Delta t$ the system interacts only with a small portion
of the bath, and being the bath of infinite dimension, the portion with
which the system is interacting changes at each time step. Effectively we couple the system with a single ancillary qubit, representing a part of the bath, evolve the full state for $\Delta t$ and then reset the bath qubit to the ground state. This last step can
be seen as changing the part of the bath connected to the system (see Fig. \ref{disegno}).

This idea corresponds to selecting the following finite representation for the bath operators
\begin{equation}
\label{finite_dB}
   \D\hat{B}_k(t)\to \hat{\sigma}_{E}^{-}\D W_k(t),
    \quad \D \hat{B}_k^{\dagger}(t) \to \hat{\sigma}_{E}^{+}\D W_k(t),
\end{equation} 
where $\hat{\sigma}_{E}^{-} = \ket{0}_{E}\bra{1}_{E}$, $\hat{\sigma}_{E}^{+} = \ket{1}_{E}\bra{0}_{E}$ and $\D W_k(t)$ are classical Wiener processes. By using the latter definitions and applying the following prescriptions
\begin{enumerate}

\item Substitute $\mathbb{E}_{\text{Q}}[\,\cdot \,]$ with $\mathbb{E}_{\text{C}}[\Tr_{\text{E}}(\hat{\rho}_{\text{in}}\,\cdot \,)]$, where $\mathbb{E}_{C}[\, \cdot \,]$ is the average over classical stochastic processes and $\hat{\rho}_{\text{in}}=|0\rangle_E \langle 0|_E$ is the ground state of the bath qubit,
\item At each time step $\Delta t$ the state is factorized as

$\ket{\Psi} = \ket{\psi}_{S}\ket{0}_{\text{E}}$,
\item The density matrix of the system has to be computed as

$\hat{\rho}_{S} = \mathbb{E}_{C}[\Tr_{\text{E}}(\ket{\Psi}\bra{\Psi})]$,
\end{enumerate}
one reproduces Eq. \eqref{mean_quantum_wiener} and the master equation of the open system in Eq. \eqref{Lindblad}. This proves that for a generic $\hat{H}_{S}$ and generic Lindbald operators a single bath qubit is enough to reproduce the open dynamics. The complete derivation of the master equation is in Appendix \ref{finite_repres_deriv}. Prescription 2 follows directly from the fact that in the finite dimensional representation the bath qubit must be reset after each time step to preserve Markovianity; the first and the third prescriptions are required to average over classical stochastic processes, thus to obtain the right result.

By assuming the finite representation \eqref{finite_dB} of the bath operators the stochastic terms in Eq. \eqref{S_k} become
\begin{equation}
\label{S_k_2}
\hat{S}_k(\Delta t) = \int_t^{t+\Delta t} \Bigl(\hat{L}_k(s) \hat{\sigma}_{E}^{+} - \hat{L}_k^{\dagger}(s) \hat{\sigma}_{E}^{-}\Bigr)\D W_k(s).
\end{equation}
For convenience we define $\hat{J}_k(s) \equiv \hat{L}_k(s) \hat{\sigma}_{E}^{+} - \hat{L}_k^{\dagger}(s) \hat{\sigma}_{E}^{-}$. The real $[\hat{S}_{k}(\Delta t)]^{\text \tiny{R}}_{ij}$ and imaginary part $[\hat{S}_{k}(\Delta t)]^{\text \tiny{I}}_{ij}$ of the entries of the operators $\hat{S}_{k}(\Delta t)$ in Eq.~\eqref{S_k_2} are It\^o integrals of deterministic functions $[\hat{S}_{k}(\Delta t)]^{\lambda}_{ij} = \int_{t}^{t+\Delta t} \D W_{k}(s)[\hat{J}_{k}(s)]^{\lambda}_{ij}$ for $\lambda = \text \tiny{R} ,\tiny{I}$, that represent Gaussian stochastic processes with means zero $\mathbb{E}\big[[\hat{S}_{k}(\Delta t)]^{\lambda}_{ij}\big] = 0$, variances $\mathbb{V}\bigl[[\hat{S}_{k}(\Delta t)]^{\lambda}_{ij}\bigr] = \int_{t}^{t+\Delta t} \D s([\hat{J}_{k}(s)]^{\lambda}_{ij})^2$ and covariances $\mathbb{E}\bigl[[\hat{S}_{k}(\Delta t)]^{\lambda}_{ij} [\hat{S}_{k}(\Delta t)]^{\lambda '}_{i'j'}\bigr] =  \int_{t}^{t+\Delta t} \D s[\hat{J}_{k}(s)]^{\lambda}_{ij}[\hat{J}_{k}(s)]^{\lambda '}_{i'j'}$.
In a quantum simulation once all the variances and covariances are computed, the stochastic processes can be sampled to get a single realization of the evolution $\hat{N}(\Delta t)$. Finally by averaging over all the realizations of the final state $\hat{\rho}_{S}(t+\Delta t) = \mathbb{E}_C[\Tr_{E}\big(\hat{N}(\Delta t)\ket{\Psi(t)}\bra{\Psi(t)}\hat{N}^{\dagger}(\Delta t)\big)]$, in accordance with the prescriptions, one obtains the same final density matrix in Eq. \eqref{rho_ng_approx}.

\section{Algorithm implementation}\label{sec:algorithm}
In Alg. \ref{alNG} we now describe the algorithmic implementation of the approach.
\begin{algorithm}[H]
\caption{}
\label{alNG}
\KwIn{A Lindblad equation with system Hamiltonian $\hat{H}_S$ and a set of $m$-local Lindblad operators; a number $\mathcal{N}_r$ of samples to be performed.}
{\bf Protocol:}
\begin{enumerate}
{\item Divide the unitary evolution of the closed system $\hat{U}(T)=e^{-\frac{i}{\hbar}\hat{H}_S T}$ in $\Delta t$ time steps as $\hat{U}(T) = \prod_{j}^{N_{\text{step}}}\hat{U}(\Delta t)$ where $T = N_{\text{step}}\Delta t$.}\;
\BlankLine
{\item Compute $\hat{N}(\Delta t)$ corresponding to $\hat{U}(\Delta t)$ by using Eq. \eqref{Q_noisy_gate} in the finite representation of Sec. \ref{sec:finite_repr}};
\BlankLine
{\item Apply $\prod_{j}^{N_{\text{step}}}\hat{N}(\Delta t)$ making sure to reset the ancilla qubit after the application of each $\hat{N}(\Delta t)$.}
\BlankLine
{\item Repeat the resulting quantum circuit of point 3 by sampling $\mathcal{N}_r$ times the stochastic processes inside each $\hat{N}(\Delta t)$.}\;
\BlankLine
\end{enumerate}
\KwOut{Final probabilities computed by taking the average over all $\mathcal{N}_r$ circuit realizations.}
\end{algorithm}

\begin{figure*}
\centering
\begin{minipage}{0.325\textwidth}
\includegraphics[width=\textwidth]{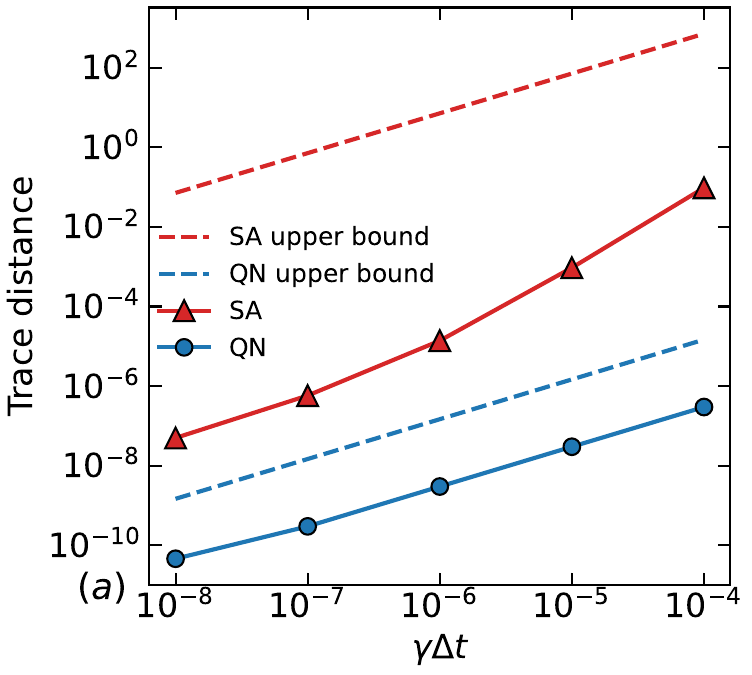}
\end{minipage}
\begin{minipage}{0.325\textwidth}
\includegraphics[width=\textwidth]{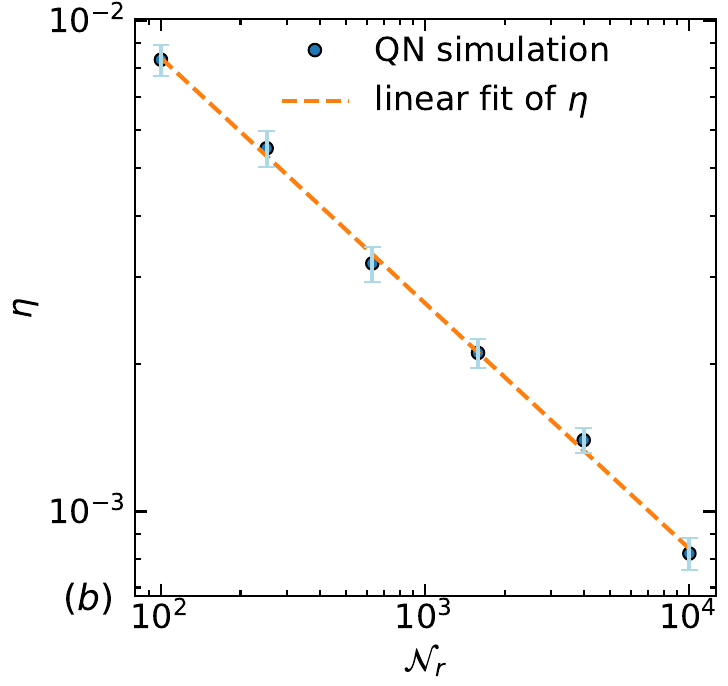}
\end{minipage}
\caption{Evolution of a single spin under external magnetic field and in contact with a thermal bath with coupling constant $\gamma = 0.1\,\text{kHz}$, for a total time $T = 30\,\mu\text{s}$. Panel (a) displays in blue the trace distance $\mathcal{T}^{(QN)}(T)$ between the exact solution of the Lindblad equation and the quantum noise (QN) approximation in Eq. \eqref{rho_ng_approx} as a function of $\gamma \Delta t$ and in red the trace distance $\mathcal{T}^{(SA)}(T)$ between the exact solution of the Lindblad equation and the standard approximation (SA). As one can see, to reach the same precision of QN, the SA needs $N^{(SA)}_{step}\sim 10^{5}$ against $N^{(QN)}_{step}\sim 10^{1}$, implying that the resulting depth for SA is four orders magnitude larger than for QN. The blue dashed line reports the theoretical upper bound for QN computed in Appendix \ref{approx_error_appendix}, while the red dashed line is the theoretical upper bound for SA, computed by using the formula in \cite{cattaneo2021collision}.
The blue dots in panel (b) are the values of $\eta = |\langle\hat{Z}\rangle-\langle\hat{Z}\rangle_{\mathcal{N}_{r}}|$ for the observable $\hat{Z}$ where $\langle\hat{Z}\rangle$ is the true expected value and $\langle\hat{Z}\rangle_{\mathcal{N}_{r}}$ is the estimate with $\mathcal{N}_{r}$ total number of realizations obtained by simulating Alg. \ref{alNG} with fixed $\Delta t = 10^{-6}$ s, $\gamma = 0.1$ kHz and for different number of realizations $\mathcal{N}_r$. Each point is the mean over $100$ independent simulations, and the  vertical bars show the standard deviations of the means. The orange dashed line is a fit for the sampling error, showing that $\eta \sim \mathcal{O}(1/\sqrt{\mathcal{N}_r})$. By taking larger $\mathcal{N}_r$ one can reach a precision consistent with the one of the rightmost point of the blue curve in (a), that corresponds to the same choice of parameters $\Delta t$ and $\gamma$; this requires a very large number of samples, but it is a consequence of the simple toy model we are considering, as explained in the main text.}
\label{trace_distance}
\end{figure*}

\begin{figure}[t!]
\centering
\includegraphics[width=0.45\textwidth]{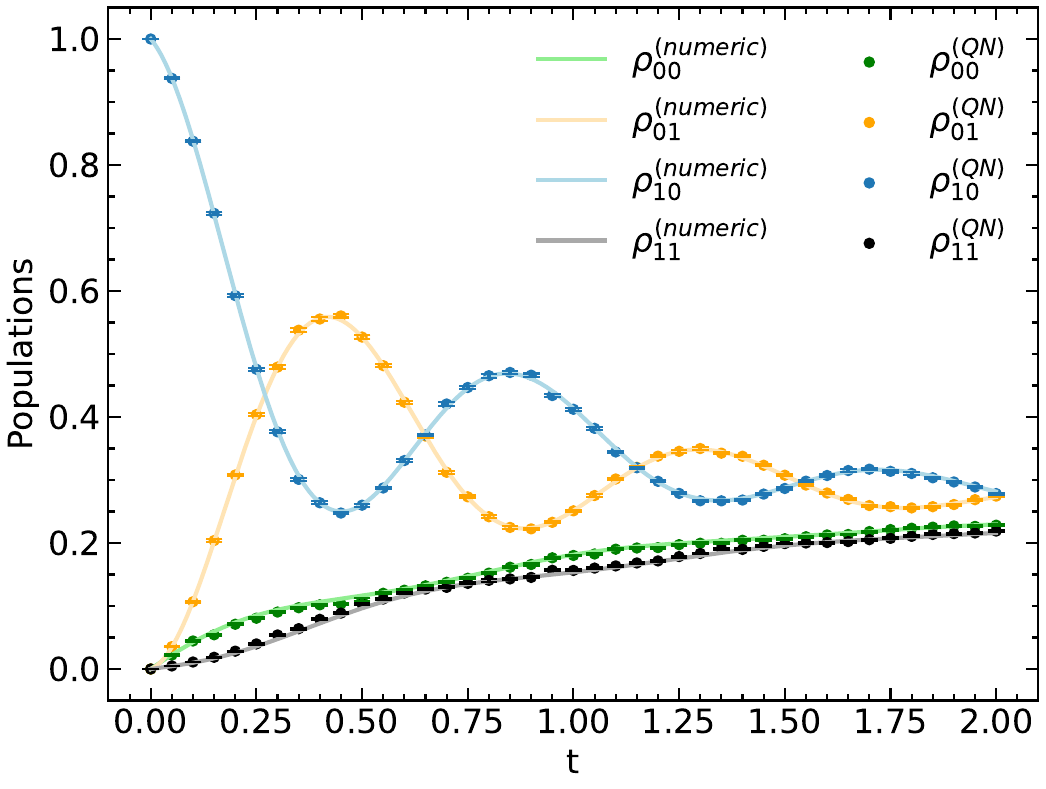}
\caption{Evolution of the diagonal entries of the density matrix describing the energy transfer between two molecules represented by two qubits. The results of our algorithm are shown in dots, while those obtained by numerically solving the Lindblad equation are shown with solid lines. The initial state of the two qubits is $\ket{10}$. All the chosen parameters are dimensionless. The values of the energy gaps are $E_0 \simeq 773.5$, $E_1 \simeq 770.3$. The interactions coupling reads $J_{0,1} = 3.2$. We chose $\Delta t = 5\times 10^{-2}$ with $N_{step} = 40$, and a number of shots $\mathcal{N}_r = 100$. The values of $\gamma_k$ are listed in Appendix \ref{coupling_constants_values}.}
\label{fig:two_qubits}
\end{figure}

We notice that when $\hat{H}_S = \sum_{\alpha=1}^{K} \hat{H}_\alpha$, with $[\hat{H}_\alpha,\hat{H}_\beta] \ne 0$, then at point 2 of Alg. \ref{alNG} one can apply the first order Trotter-Suzuki product formula $\hat{U}(s-t)\simeq\prod_{\alpha=1}^K e^{-\frac{i}{\hbar}\hat{H}_{\alpha} (s-t)}$ \cite{suzuki1985decomposition}. Since for relevant applications each $\hat{H}_\alpha$ is $m$-local, then by plugging the latter expression in Eq. \eqref{Q_noisy_gate}, all the resulting terms have a locality that depends on $m$ but not on $n$ (see Appendix \ref{locality}).

Given the analysis of the perturbative approximation of Sec. \ref{sec:approx_solution}, in Appendix \ref{total_error_deltat} we prove that each $\hat{N}(\Delta t)$ contributes with an error $\varepsilon \leq \varepsilon_T + \varepsilon_p$, where $\varepsilon_p$ is  defined in Eq. \eqref{e_p} and $\varepsilon_T$ is the error due to the first order Trotter-Suzuki product formula, upper bounded by
\begin{equation}
\label{epsilon_T}
\varepsilon_T \leq \Bigl(KJ\max_{\alpha,j}\norm{\hat{h}_{\alpha,j}}_{\infty} \omega\Delta t\Bigr)^2,
\end{equation}
where $\omega$ is the strength of the Hamiltonian $\hat{H}_S$, assuming for simplicity that the frequencies of each $\hat{H}_\alpha$ have the same order of magnitude and $\hat{H}_\alpha = \omega\sum_{j=1}^J, \hat{h}_{\alpha,j}$ where $J$ is a constant whose value depends on the system Hamiltonian under study and $\hat{h}_{\alpha,j}$ are generic $m$-local operators.
Thus the global approximation error for the total time $T$ is $\varepsilon_{global} = N_{step}\,\varepsilon$ which by using Eqs. \eqref{bound_e_p} and \eqref{epsilon_T} is upper bounded by
\begin{equation}
\label{e_global}
\begin{aligned}
\varepsilon_{global} &\leq \frac{T^2K^2}{N_{step}}\biggl(J^2\omega^2\max_{\alpha,j}\norm{\hat{h}_{\alpha,j}}_{\infty}^2\\
&\quad+(2^{2m}-1)^2\gamma^2\max_{k,j}\norm{\hat{L}_{k,j}}^{4}_{\infty}\biggr).
\end{aligned}
\end{equation}
This bound is polynomial in the system qubits number $n$ as $K \sim \mathcal{O}(n^m)$.

Additionally, we show in Appendix \ref{higher_trotter} that the algorithm works also for higher order of the Trotter-Suzuki product formula, improving the upper bound on $\varepsilon_T$ in Eq. \eqref{epsilon_T}.

\section{Resources estimation}

In order to simulate the dynamics up to time $T$ making an error not greater than $\varepsilon_{global}$, our algorithm needs
\begin{equation}
\label{stima_numero_gates}
\begin{aligned}
\text{N}_G &\propto \biggl\lceil \biggl(KJ + K (2^{2m}-1) +1 \biggr) \frac{T^2 K^2(\gamma^2+\omega^2)}{\varepsilon_{global}}\biggr\rceil\\
&\sim \mathcal{O}\Biggl(\biggl\lceil  \frac{n^{4m} T^2 (\gamma^2+\omega^2)}{\varepsilon_{global}}\biggr\rceil\Biggr)
\end{aligned}
\end{equation}
total number of gates. In Eq. \eqref{stima_numero_gates} $KJ$ is the number of gates to implement $\hat{U}(\Delta t)$, that is polynomial in $n$ for $m$-local Hamiltonians. The term $K (2^{2m}-1)$ counts the number of gates needed to implement the product in Eq. \eqref{Q_noisy_gate}, where $K$ scales polynomially with $n$. The term $+1$ counts the reset of the ancillary qubit. Thus, $\text{N}_G = \text{poly}(n, \gamma T,\omega T, 1/\varepsilon_{global})$ and therefore our algorithm is efficiently simulatable on a quantum computer \cite{kliesch2011dissipative}.

Given a fixed accuracy, our approximation (see Eqs. \eqref{Q_noisy_gate} and \eqref{rho_ng_approx}) allows to implement a number of trotter steps which is smaller than those required in other approaches \cite{cattaneo2021collision, cattaneo2023quantum,pocrnic2024quantum}, implying a reduction of the circuit depth. See Appendix \ref{domain_application_appendix} for further details.

In general our ancilla overhead is constant and is always equal to one, thus of order $\mathcal{O}(1)$, as shown above, while many of the state of the art implementations \cite{sweke2015universal,de2021quantum,cattaneo2021collision, cattaneo2023quantum, cleve2016efficient,kamakari2022digital} have an ancilla overhead proportional to the number of Lindblad operators thus in general of order $\mathcal{O}(2^{2n}-1$) where $n$ is the number of system qubits or, assuming a $m$-locality condition, of order $\mathcal{O}(n^m)$, or if the implementation is based on the vectorization of the density matrix, of order $\mathcal{O}(n)$. Other methods, such as the one presented in \cite{kliesch2011dissipative}, are able to use the same set of $2^{2m}$ ancillary qubits for every $m$-local term in the Lindbladian, i.e. an asymptotic overhead of order $\mathcal{O}(1)$. However these methods need a number of ancillary qubits that is exponential in the $m$-locality, while our algorithm always need a single ancilla regardless of both $n$ and $m$. Already in the most simple case of $m$ = 2, the algorithms in \cite{kliesch2011dissipative} need 16 ancillas versus 1 ancilla of our method. This is a relevant achievement especially in the NISQ era.

Finally, our approach needs to evaluate multiple circuits due to the sampling of the stochastic processes and this leads to a 
sampling error on the evaluation of expectation values of an $m$-local observable $\hat{O}$ given by $\eta \sim \mathcal{O}(1/\sqrt{\mathcal{N}_r})$ with $\mathcal{N}_{r}$ the total number of samples of the classical stochastic processes. Thus $\eta$ does not scale asymptotically with the size of the system. We give further details on $\eta$ in Appendix \ref{sampling_error_appendix}.

\section{Proof of concept}
We now provide a proof of concept of the effectiveness of the approach. We consider a toy model dynamics of a single spin under external magnetic field with system Hamiltonian $H_{S} = \frac{\hbar\Omega}{2}\hat{X}$ with $\Omega = \pi/6 \,\text{MHz}$ and Lindblad operators $\hat{L}_{1} = \hat{\sigma}^{+}$, $\hat{L}_{2} = \hat{\sigma}^{-}$ and $\hat{L}_{3} = \hat{Z}$ driven by the same parameter $\gamma =0.1 
 \,\text{kHz}$. The isolated dynamics of the system can be implemented by a single-qubit x-rotation $\hat{R}_{x}(\theta)$ in time steps $\Delta t$  where $\theta = \Omega \Delta t$.
First we compute the trace distances \cite{nielsen2000quantum} $\mathcal{T}^{(QN)} = \mathcal{T}(\hat{\chi}(T),\hat{\rho}^{(QN)}(T))$ and $\mathcal{T}^{(SA)} = \mathcal{T}(\hat{\chi}(T),\hat{\rho}^{(SA)}(T))$ where $\hat{\chi}(T)$ is the analytic solution of the Lindblad equation, $\hat{\rho}^{(QN)}(T)$ is obtained with our approximation in Eq.\eqref{rho_ng_approx} that from now on we call quantum noise (QN) approximation, $\hat{\rho}^{(SA)}(T)$ is obtained with the approximation $\hat{\rho}_{S}(t+\Delta t) =\hat{\rho}_{S}(t)+\bigl(-\frac{i}{\hbar}[\hat{H}_{S},\hat{\rho}_{S}(t)]+\mathcal{D}\hat{\rho}_{S}(t)\bigr)\Delta t$, that from now on we call standard approximation (SA). We fix the total evolution time $T = 30\,\mu\text{s}$ for different values of $\Delta t$. In Fig.\ref{trace_distance}(a) we plot the results. The red and blue curves are respectively the trends of $\mathcal{T}^{(SA)}$ and $\mathcal{T}^{(QN)}$ for different values of $\gamma \Delta t$. The theoretical upper bound for QN computed in Appendix \ref{approx_error_appendix} is reported as a blue dashed line and the theoretical upper bound for SA as a red dashed line, computed by using the formula in \cite{cattaneo2021collision}. Notably, for $\gamma \Delta t \sim 10^{-4}$, $\mathcal{T}^{(QN)} \sim 10^{-7}$ while $\mathcal{T}^{(SA)} \sim 10^{-1}$. To reach the same precision of QN the SA needs $\gamma \Delta t \sim 10^{-8}$, meaning that $\Delta t$ has to be reduced by four orders of magnitude and consequently $N^{(SA)}_{step}\sim 10^{5}$ against $N^{(QN)}_{step}\sim 10^{1}$.

The blue dots in Fig. \ref{trace_distance} (b) are the values of $\eta = |\langle\hat{Z}\rangle-\langle\hat{Z}\rangle_{\mathcal{N}_{r}}|$ for the observable $\hat{Z}$ obtained by simulating Alg. \ref{alNG} with fixed $\Delta t = 10^{-6}$ s, $\gamma = 0.1$ kHz and for different number of realizations $\mathcal{N}_r$. Each point is the mean over $100$ independent simulations, vertical bars are the standard deviations of the means. The orange dashed line is a linear fit in log-log scale of the data, giving angular coefficient $\simeq -1/2$, showing that the behaviour of the sampling error is $\eta \sim \mathcal{O}(1/\sqrt{\mathcal{N}_r})$.  By taking larger $\mathcal{N}_r$ one can reach a precision consistent with the one of the rightmost point of the blue curve in Fig. \ref{trace_distance} (a), that corresponds to the same choice of parameters $\Delta t$ and $\gamma$. This requires a huge number of samples, but it is a consequence of the simple toy model of a single system qubit interacting with an environment, for which the approximation error is very small. When considering systems with high number of qubits, since the approximation error scales with $\sim n^{2m}$, a limited number of realizations $\mathcal{N}_r$ is required to reach the accuracy of the approximation error.

Finally we simulated classically Alg. \ref{alNG} applied to the problem of the energy transfer dynamics in molecular systems. In particular we considered the following Hamiltonian for a linear chain of two molecules
\begin{equation}
\hat{H}_S = -\sum_{k=0}^1 \frac{E_k}{2}\hat{Z}_k + \frac{J_{0,1}}{2}\Bigl(\hat{X}_0\hat{X}_1 + \hat{Y}_0\hat{Y}_1\Bigr),
\end{equation}
where each molecule has only two energy levels encoded in the qubits computational states. All the chosen parameters are dimensionless. The values of the energy gaps are $E_0 \simeq 773.5$, $E_1 \simeq 770.3$. The interactions coupling reads $J_{0,1} = 3.2$. The dissipator of the Lindblad equation we used is
\begin{equation}
\mathcal{D}\hat{\rho}_S = \sum_{k=0}^{15}\gamma_k \Bigl(\hat{\mathcal{P}}_k\hat{\rho}_S\hat{\mathcal{P}}_k - \hat{\rho}_S\Bigr),
\end{equation}
where $\hat{\mathcal{P}}_k$ are two-qubits Pauli strings and the values of $\gamma_k$ are listed in Appendix \ref{coupling_constants_values}. The initial state of the two qubits is $\ket{10}$. We chose $\Delta t = 5\times 10^{-2}$ with $N_{step} = 40$, and a number of shots $\mathcal{N}_r = 100$. In Fig. \ref{fig:two_qubits} we show the evolution of the diagonal entries of the density matrix. Solid lines are numeric solutions while dots are obtained with our algorithm. The good agreement between the two confirm the validity of the method.

\section{Conclusions}
We have presented an algorithm based on the quantum noise formalism tailored for the efficient simulation of open quantum systems on quantum devices. Our approach allows to use a single ancilla qubit of the bath independently of the system size marking a substantial reduction in the total number of circuit qubits. Moreover our approximation, for small environment coupling constants, permits to reduce the total number of steps while maintaining the desired accuracy. For bigger values of the environment coupling constants, our approach still works and converges to the standard first order solution in $\Delta t$. Moreover, in principle, the accuracy can be improved by expanding the approximation to higher orders, see Appendices \ref{perturbative_solution_appendix} and \ref{higher_trotter}.

As future work we expect to apply the algorithm on real quantum computers paired with error mitigation techniques to reduce the impact of the inherent noise of the devices. We will also explore the possibility of introducing non-Markovian effects by relaxing the prescription of resetting the bath qubit after each gate in the time step.

\begin{acknowledgments}
\noindent
G.D.B. and M.V. thank L.L. Viteritti for useful discussions. T.F., G.D.B. and M.V. acknowledge the financial support from University of Trieste and INFN; S.D. acknowledges support from the  Marie Sklodowska Curie Action through the UK Horizon Europe guarantee administered by UKRI, and from INFN. A.B. acknowledges financial support from the EIC Pathfinder project QuCoM (GA no. 101046973), the PNRR PE National Quantum Science and Technology Institute (PE0000023), the University of Trieste and INFN.
\end{acknowledgments}

\appendix

\onecolumngrid

\section{The quantum noise formalism}
\label{appendix_A}
Our approach is based on the quantum noise formalism \cite{gardiner2004quantum,wiseman2009quantum, zoller1997quantum}, that here we briefly introduce. We consider the following Hamiltonian $\hat{H} = \hat{H}_{\text{S}} + \hat{H}_{\text{E}} + \hat{H}_{\text{SE}}$ where $\hat{H}_{\text{S}}$ is the system Hamiltonian while $\hat{H}_{\text{E}}$ and $\hat{H}_{\text{SE}}$ are, respectively, the environment Hamiltonian and the system-environment interaction Hamiltonian, that read
\begin{align}
\hat{H}_{\text{E}}& = \hbar \sum_k\int \D \omega \omega \hat{b}_k^{\dagger}(\omega)\hat{b}_k(\omega),\\
\hat{H}_{\text{SE}} &= i\hbar \sum_k\int \D \omega \chi_k(\omega)\Bigl(\hat{b}_k^{\dagger}(\omega)\hat{L}_k-\hat{L}_k^{\dagger}\hat{b}_k(\omega)\Bigr),
\end{align}
where $\hat{b}_k(\omega)$ is the annihilation operator of the $k$-th thermal bath, $\omega$ is the frequency of the mode and $\hat{L}_k$ is an arbitrary system operator. By using the Hamiltonian  $\hat{H}$ one can solve the Heisenberg equation for $\hat{b}_k(\omega)$ and substitute the solution in the Heisenberg equation for an arbitrary operator $\hat{O}$. In the Markov approximation, namely when $\chi_k(\omega) = \sqrt{\gamma_k/(2\pi)}$, the Heisenberg equation for the operator $\hat{O}$ reads \cite{gardiner2004quantum}:
\begin{equation}
\label{langevin_app}
\begin{aligned}
\frac{\D}{\D t}\hat{O} = \frac{i}{\hbar}[\hat{H}_{\text{S}},\hat{O}]+\sum_k\gamma_k\Bigl(\hat{L}_k^{\dagger}\hat{O}\hat{L}_k-\frac{1}{2}\{\hat{L}_k^{\dagger}\hat{L}_k,\hat{O}\}\Bigr)+\sum_k\sqrt{\gamma_k}\Bigl(\hat{b}_{\text{in},k}^{\dagger}(t)[\hat{O},\hat{L}_k]+\hat{b}_{\text{in},k}(t)[\hat{L}_k^{\dagger},\hat{O}]\Bigr),
\end{aligned}
\end{equation}
where $\hat{b}_{\text{in},k}(t) = \frac{1}{\sqrt{2 \pi}} \int \D \omega e^{-i\omega(t-t_0)}\hat{b}_k(t_0,\omega)$.

Now we consider the ground state $\ket{0_\omega}_{\text{E}}$ of the thermal baths and define $\hat{\rho}_{\text{in}} = \ket{0_\omega}_{\text{E}}\bra{0_\omega}_{\text{E}}$. In this way by defining $\mathbb{E}_{\tiny\text{Q}}[\,\cdot\,] \equiv \Tr_{\text{E}}(\hat{\rho}_{\text{in}}\,\cdot\,)$ the following relation holds
$\mathbb{E}_{\text{Q}}[\hat{b}_k(\omega)\hat{b}_j^{\dagger}(\omega')] = \Tr_{\text{E}}(\hat{\rho}_{\text{in}}\hat{b}_k(\omega)\hat{b}_j^{\dagger}(\omega')) = \delta_{kj}\delta(\omega-\omega')$
and the mean values of the other combinations of two thermal baths operators are zero. We notice that if we define $\D \hat{B}_k(t) = \hat{b}_{\text{in},k}(t)\D t$, then by using this latter operator the only non zero mean value is
\begin{equation}
\label{mean_quantum_wiener_app}
\mathbb{E}_{\text{Q}}\Bigl[\D \hat{B}_k(t)\D\hat{B}_j^{\dagger}(t)\Bigr] = \delta_{kj}\D t.
\end{equation}
 Thus in force of Eq. \eqref{mean_quantum_wiener} the operator $\hat{B}_k(t)$ can be interpreted as the quantum generalization of a Wiener process \cite{gardiner1985handbook, jacobs2010stochastic} and from Eq. \eqref{langevin_app} one can build the following quantum stochastic differential equation (QSDE) for the state vector of the system plus environment $\ket{\Psi(t)}$
\begin{equation}
 \label{QSDE_app}
\begin{aligned}
   \D\ket{\Psi(t)}= \Bigl[ -\frac{i}{\hbar}\hat{H}_{\text{S}}\D t + \sum_k\sqrt{\gamma_k}\left(\hat{L}_k\,\D \hat{B}_k^{\dagger}(t) - \hat{L}_k^{\dagger} \,\D \hat{B}_k(t) \right) -\sum_k\frac{\gamma_k}{2} \hat{L}_k^{\dagger} \hat{L}_k dt \Bigr]\ket{\Psi(t)}.
\end{aligned} 
\end{equation}
Finally from Eq. \eqref{QSDE_app} we can compute the master equation for the density matrix of the open quantum system (see Appendix \ref{standard_deriv}) in Eq. \eqref{Lindblad} that is in the Lindblad form \cite{gorini1976completely,lindblad1976generators,breuer2002theory}, where $\hat{\rho}_{S}$ is such that $\hat{\rho}_{S}(t) = \Tr_{\text{E}}(\ket{\Psi(t)}\bra{\Psi(t)})$.

\section{Derivation of the master equation of the system}
We show how to recover the Lindblad equation for the system density matrix $\hat{\rho}_{S}$, first following the standard derivation and then in the case of finite representation of the bath operators. 

\subsection{Standard derivation}\label{standard_deriv}
 We start from the quantum stochastic differential equation (QSDE) in the main text. For simplicity we consider to have a single Lindblad operator $\hat{L}$ with coefficient $\gamma$, the generalization to more Lindblad operators being straightforward.
Starting from $\hat{\rho}_{S} = \Tr_{\text{E}}(\ket{\Psi}\bra{\Psi})$, we differentiate on both sides to get
\begin{equation}
   \D \hat{\rho}_{S} = \Tr_{\text{E}}\bigl(\D \ket{\Psi}\bra{\Psi} + \ket{\Psi}\D\bra{\Psi} + \D \ket{\Psi}\D \bra{\Psi}\bigr).
\end{equation}
By using the expression for $\D \ket{\Psi}$, its conjugate for $\D\bra{\Psi}$, and neglecting terms of order $\mathcal{O}(\D t^2)$ one gets
\begin{equation}\label{derivation_lindblad_dB}
\begin{split}
\D \hat{\rho}_{S} = &-\frac{i}{\hbar}[\hat{H}_{\text{S}},\hat{\rho}_{S}]\D t-\frac{\gamma}{2}\{\hat{L}^{\dagger}\hat{L},\hat{\rho}_{S}\}\D t +\sqrt{\gamma}\hat{L} \Tr_{\text{E}}\bigl(\D\hat{B}^{\dagger}\ket{\Psi}\bra{\Psi}\bigr) -\sqrt{\gamma}\hat{L}^{\dagger} \Tr_{\text{E}}\bigl(\D \hat{B}\ket{\Psi}\bra{\Psi}\bigr)+\sqrt{\gamma} \Tr_{\text{E}}\bigl(\ket{\Psi}\bra{\Psi}\D \hat{B}\bigr)\hat{L}^{\dagger}\\ 
&- \sqrt{\gamma}\Tr_{\text{E}}\bigl(\ket{\Psi}\bra{\Psi}\D\hat{B}^{\dagger}\bigr) \hat{L}+\gamma\hat{L} \Tr_{\text{E}}\bigl(\D\hat{B}^{\dagger}\ket{\Psi}\bra{\Psi}\D\hat{B}\bigr)\hat{L}^{\dagger} -\gamma\hat{L} \Tr_{\text{E}}\bigl(\D\hat{B}^{\dagger}\ket{\Psi}\bra{\Psi}\D\hat{B}^{\dagger}\bigr)\hat{L}-\gamma\hat{L}^{\dagger}\Tr_{\text{E}}\bigl(\D\hat{B}\ket{\Psi}\bra{\Psi}\D\hat{B}\bigr)\hat{L}^{\dagger}  \\
&+ \gamma\hat{L}^{\dagger} \Tr_{\text{E}}\bigl(\D\hat{B}\ket{\Psi}\bra{\Psi}\D\hat{B}^{\dagger}\bigr)\hat{L}= \\
&=-\frac{i}{\hbar}[\hat{H}_{\text{S}},\hat{\rho}_{S}]\D t-\frac{\gamma}{2}\{\hat{L}^{\dagger}\hat{L},\hat{\rho}_{S}\}\D t+\gamma\hat{L}\hat{\rho}_{S}\hat{L}^\dagger\D t
\end{split}
\end{equation}
where we used the fact that $\D\hat{B}\ket{\Psi} = 0$ at all times \cite{zoller1997quantum} together with the ciclicity of the trace, which implies that the only term surviving is the second term in the second line, leading to the master equation in the Lindblad form. We mention that this derivation is valid for a generic state $\ket{\Psi}$ of the system-environment as far as this state is evolved from an initial state which is factorized with the bath in the ground state i.e. $\ket{\psi(0)}_S\ket{0_\omega}_E$. 

\subsection{Derivation in the case of finite representation of the bath operators}\label{finite_repres_deriv}
We now substite the finite representation of the bath operators in the main text inside the first line of Eq.\eqref{derivation_lindblad_dB} to get
\begin{equation}\label{derivation_lindblad_dW}
\begin{split}
\D \hat{\rho}_{S} = &\mathbb{E}_{C}\Bigg[-\frac{i}{\hbar}[\hat{H}_{\text{S}},\hat{\rho}_{S}]\D t-\frac{\gamma}{2}\{\hat{L}^{\dagger}\hat{L},\hat{\rho}_{S}\}\D t+\sqrt{\gamma}\hat{L}\D W\Tr_{\text{E}}\bigl(\hat{\sigma}^{+}_{E}\ket{\Psi}\bra{\Psi}\bigr) -\sqrt{\gamma}\hat{L}^{\dagger} \D W\Tr_{\text{E}}\bigl(\hat{\sigma}^{-}_{E}\ket{\Psi}\bra{\Psi}\bigr)\\
&+\sqrt{\gamma} \Tr_{\text{E}}\bigl(\ket{\Psi}\bra{\Psi}\hat{\sigma}^{-}_{E}\bigr)\hat{L}^{\dagger}\D W - \sqrt{\gamma}\Tr_{\text{E}}\bigl(\ket{\Psi}\bra{\Psi}\hat{\sigma}^{+}_{E}\bigr) \hat{L}\D W +\gamma\hat{L} \Tr_{\text{E}}\bigl(\ket{\Psi}\bra{\Psi}\hat{\sigma}^{-}_{E}\hat{\sigma}^{+}_{E}\bigr)\hat{L}^{\dagger}\D t \\
&-\gamma\hat{L} \Tr_{\text{E}}\bigl(\ket{\Psi}\bra{\Psi}\hat{\sigma}^{+}_{E}\hat{\sigma}^{+}_{E}\bigr)\hat{L}\D t-\gamma\hat{L}^{\dagger}\Tr_{\text{E}}\bigl(\ket{\Psi}\bra{\Psi}\hat{\sigma}^{-}_{E}\hat{\sigma}^{-}_{E}\bigr)\hat{L}^{\dagger}\D t  + \gamma\hat{L}^{\dagger} \Tr_{\text{E}}\bigl(\ket{\Psi}\bra{\Psi}\hat{\sigma}^{+}_{E}\hat{\sigma}^{-}_{E}\bigr)\hat{L}\D t\Bigg].
\end{split}
\end{equation}
By using the fact that $(\hat{\sigma}_{E}^{\pm})^2 = 0$, $\hat{P}_0 = \ket{0}_{E}\bra{0}_{E} = \hat{\sigma}^{-}_{E}\hat{\sigma}^{+}_{E}$ and $\hat{P}_1 = \ket{1}_{E}\bra{1}_{E}  = \hat{\sigma}^{+}_{E}\hat{\sigma}^{-}_{E}$ and that $\mathbb{E}_{C}[dW]=0$, Eq. \eqref{derivation_lindblad_dW} becomes
\begin{equation}\label{derivation_lindblad_dW_1}
\begin{split}
\D \hat{\rho}_{S} =& -\frac{i}{\hbar}[\hat{H}_{\text{S}},\hat{\rho}_{S}]\D t-\frac{\gamma}{2}\{\hat{L}^{\dagger}\hat{L},\hat{\rho}_{S}\}\D t+\gamma\hat{L} \Tr_{\text{E}}\bigl(\ket{\Psi}\bra{\Psi}\hat{P}_0\bigr)\hat{L}^{\dagger}\D t + \gamma\hat{L}^{\dagger} \Tr_{\text{E}}\bigl(\ket{\Psi}\bra{\Psi}\hat{P}_1\bigr)\hat{L}\D t.
\end{split}
\end{equation}
The Lindblad master equation is recovered by using the prescriptions of the main text. Since $\ket{\Psi} = \ket{\psi}_{S}\ket{0}_{\text{E}}$ the $\hat{P}_1$ term give a zero contribution and $\Tr_{\text{E}}\bigl(\ket{\Psi}\bra{\Psi}\hat{P}_{0}\bigr)=\ket{\psi}_{S}\bra{\psi}_{S}\bra{0}_{E}\hat{P}_{0}\ket{0}_{E}=\ket{\psi}_{S}\bra{\psi}_{S}$, recovering the Lindblad equation.

\section{Perturbative solution of the QSDE}\label{perturbative_solution_appendix}
To derive the approximate solution of the QSDE in the main text we use a perturbative method known as \textit{small noise expansion} \cite{gardiner1985handbook}. The following derivation is performed by using the generic expression for $\D \hat{B}_t$ and $\D \hat{B}_t^{\dagger}$, but the same results apply when substituting their finite expression and the prescriptions introduced in the main text. For simplicity, let us consider the QSDE with one single Lindblad operator,
 \begin{equation}\label{perturbeq}
    \D\ket{\Psi_t} =\bigg[-\frac{i}{\hbar}\hat{H} \D t + \epsilon \bigl(\hat{L} \D \hat{B}_t^{\dagger}-\hat{L}^{\dagger} \D \hat{B}_t\bigr) - \frac{\epsilon^2}{2} \hat{L}^\dag\hat{L} \D t \bigg]\ket{\Psi_t},
\end{equation}
where $\epsilon = \sqrt{\gamma}$. The generalization to $2^{2n}-1$ Lindblad operators again being straightforward, and let us set the following perturbative expansion:
\begin{equation}
    \ket{\Psi_s}= \ket{\Psi_s^0}+\epsilon\ket{\Psi_s^1}+\epsilon^2\ket{\Psi_s^2}+\dots
\end{equation}
Substituting the latter into Eq.(\ref{perturbeq}) and equating terms with the same power of $\epsilon$, up to second order we have the following system of QSDEs:
\begin{equation}
\begin{aligned}
&\D \ket{\Psi_t^0} = -\frac{i}{\hbar}\hat{H} \ket{\Psi_t^0}\D t \nonumber\\
& \D \ket{\Psi_t^1} = -\frac{i}{\hbar}\hat{H} \ket{\Psi_t^1}\D t +\bigl(\hat{L}\D \hat{B}_t^{\dagger}-\hat{L}^{\dagger}\D\hat{B}_t\bigr)\ket{\Psi_t^0}\nonumber\\
&\D \ket{\Psi_t^2} = -\frac{i}{\hbar}\hat{H}\ket{\Psi_t^2}\D t +\bigl(\hat{L}\D \hat{B}_t^{\dagger}-\hat{L}^{\dagger}\D\hat{B}_t\bigr)\ket{\Psi_t^1}-\frac{1}{2}\hat{L}^{\dagger}\hat{L}\ket{\Psi_t^0}\D s,
\end{aligned}
\end{equation}
which has to be solved with the initial conditions $\ket{\Psi_0^0}=\ket{\Psi_0}$ and $\ket{\Psi_0^j}=0$ for $j>0$ . The zero-th order differential equation is the deterministic equation given by the Hamiltonian evolution alone, hence its solution is simply $\ket{\Psi_t^0}=\hat{U}_t\ket{\Psi_0}$. The solution of the first order QSDE is: 
\begin{equation}
\ket{\Psi_t^1}=\hat{U}_t \hat{S}_t\ket{\Psi_0},    
\end{equation}
 where we introduced $\hat{S}_{t}:=\int_0^t (\hat{L}_s\D \hat{B}_s^{\dagger}-\hat{L}_s^\dagger\D \hat{B}_s)$ with $\hat{L}_s=\hat{U}_s^\dagger\hat{L}\hat{U}_s$. Last, the solution to the second order QSDE is 
\begin{equation}\label{secondordersol}
    \ket{\Psi_t^2}=-\hat{U}_t\int_0^t\Big[\frac{1}{2}\hat{L}_s^\dagger\hat{L}_s\D s-(\hat{L}_s\D \hat{B}_s^{\dagger}-\hat{L}_s^{\dagger}\D \hat{B}_s)\hat{S}_s\  \Big]\ket{\Psi_0}.
\end{equation}
Then, the solution at order $\epsilon^2$ is given by $\ket{\Psi_t}=\hat{N}\ket{\Psi_0}+\mathcal{O}(\epsilon^3)$, where the evolution operator is $\hat{N}=\hat{U}\hat{N}'$, with
\begin{equation}
\begin{aligned}
\label{evolutionnn}
\hat{N}'= \mathbb{1}+\epsilon\hat{S}_t-\frac{\epsilon^2}{2}\int_0^t \hat{L}_s^\dagger\hat{L}_s\D s+\epsilon^2\int_0^t\bigl(\hat{L}_s\D\hat{B}_s^\dagger-\hat{L}_s^\dagger\D \hat{B}_s\bigr)\hat{S}_s.
\end{aligned}
\end{equation}
In order to express the solution in the form given in the main text, we make use of the following equality obtained by using the It\^o rule \cite{gardiner2004quantum,gardiner1985handbook}
\begin{equation}
\int_0^t\bigl(\hat{L}_s\D\hat{B}_s^\dagger-\hat{L}_s^\dagger\D \hat{B}_s\bigr)\hat{S}_s = \frac{1}{2}\bigg[\hat{S}^2_t+\int_0^t \D s \hat{L}_s^\dagger\hat{L}_s -\hat{C}_t\bigg],
\end{equation}
where $\hat{C}_t = \frac{1}{2}\int_0^t \bigl[\hat{S}_s,\hat{L}_s\D\hat{B}_s^\dagger-\hat{L}_s^\dagger\D \hat{B}_s\bigr] $. By substituting this expression into Eq.\eqref{evolutionnn}, we have to second order:
\begin{equation}
\begin{split}
&\hat{N}'  =  \mathbb{1}+\epsilon\hat{S}_t +\frac{\epsilon^2}{2}\hat{S}_t^2 -\epsilon^2\hat{C}_t =  e^{\epsilon\hat{S}_t-\epsilon^2 \hat{C}_t} + \mathcal{O}(\epsilon^3).
\end{split}    
\end{equation}
We can neglect the term $\epsilon^2\hat{C}_t$ which in principle contributes to order $\epsilon^2$; this is legitimate because it is a nested It\^o integral of non anticipating functions \cite{gardiner2004quantum,gardiner1985handbook}, and hence its stochastic average is 0. In this way we obtain the same expression in Eq. \eqref{Q_noisy_gate}.

\section{Proof of Eq. \eqref{rho_ng_approx}}
\label{proof_eq_7}
Here we prove that
\begin{equation}
\hat{\rho}_{S}(t+\Delta t) = \Tr_{E}(\hat{N}(\Delta t)\ket{\Psi(t)}\bra{\Psi(t)}\hat{N}^{\dagger}(\Delta t))\simeq\hat{U}(\Delta t)\biggl(\hat{\rho}_{S}(t)+\int_{t}^{t+\Delta t} \D s\mathcal{D}(s)\hat{\rho}_{S}(t)\biggr)\hat{U}^{\dagger}(\Delta t)
\end{equation}
where $\hat{N}(\Delta t)   \simeq  \hat{U}(\Delta t) \prod_k e^{\sqrt{\gamma_k}\hat{S}_k(\Delta t)}$.  In particular when $\gamma_k \Delta t$ are small given
\begin{equation}
\label{trick}
e^{\sum_k\sqrt{\gamma_{k}}\hat{S}_{k}(\Delta t)} \simeq \prod_k e^{\sqrt{\gamma_{k}}\hat{S}_{k}(\Delta t)}.
\end{equation}
 we prove that the right hand side reproduces Eq. \eqref{rho_ng_approx}.
\begin{equation}
\begin{aligned}
&\mathbb{E}_{C}\Bigl[\Tr_{E}\Bigl(\prod_ke^{\sqrt{\gamma_{k}}\hat{S}_{k}(\Delta t)}\hat{\rho}_S(0)\ket{0}\bra{0}\prod_{j}e^{\sqrt{\gamma_{j}}\hat{S}_{j}^{\dagger}(\Delta t)}\Bigr)\Bigr] = \\
&\mathbb{E}_{C}\bigl[\Tr_{E}\Bigl(\prod_k\bigl(\mathbb{1}+\sqrt{\gamma_{k}}\hat{S}_{k}(\Delta t)+\frac{1}{2}\gamma_{k}\hat{S}_{k}^2(\Delta t)\bigr)\hat{\rho}_S(0)\ket{0}\bra{0}\prod_j\bigl(\mathbb{1}+\sqrt{\gamma_{j}}\hat{S}_{j}^{\dagger}(\Delta t)+\frac{1}{2}\gamma_{j}\hat{S}_{j}^{\dagger 2}(\Delta t)\bigr)\Bigr)\Bigr]= \\
&\mathbb{E}_{C}\bigl[\Tr_{E}\Bigl(\bigl(\mathbb{1}+\sum_k\sqrt{\gamma_{k}}\hat{S}_{k}(\Delta t)+\frac{1}{2}\sum_{k}\gamma_{k}\hat{S}_{k}^{2}(\Delta t)+\sum_{k<k'}\sqrt{\gamma_{k}}\sqrt{\gamma_{k'}}\hat{S}_{k}(\Delta t)\hat{S}_{k'}(\Delta t)\bigr)
\hat{\rho}_S(0)\ket{0}\bra{0}\\
&(\mathbb{1}+\sum_j\sqrt{\gamma_{j}}\hat{S}_{j}^{\dagger}(\Delta t)+\frac{1}{2}\sum_{j}\gamma_{j}\hat{S}_{j}^{2\dagger}(\Delta t)+\sum_{j<j'}\sqrt{\gamma_{j}}\sqrt{\gamma_{j'}}\hat{S}_{j}^{\dagger}(\Delta t)\hat{S}_{j'}^{\dagger}(\Delta t))\Bigr)\Bigr]=\\
&\hat{\rho}_{S}(t)+\int_{t}^{t+\Delta t} \D s\mathcal{D}(s)\hat{\rho}_{S}(t)
\end{aligned},
\end{equation}

where $\mathcal{D}(s)\hat{\rho}_{S}(t) = \sum_k\gamma_k\bigl(\hat{L}_k(s)\hat{\rho}_{S}(t)\hat{L}_k^{\dagger}(s)-\frac{1}{2}\bigl\{\hat{L}_k^{\dagger}(s)\hat{L}_k(s),\hat{\rho}_{S}(t)\bigr\}\bigr)$ and we use the fact that
\begin{equation}
\begin{aligned}
&\mathbb{E}_{C}\bigl[\Tr_{E}\Bigl(\sqrt{\gamma_k}\sqrt{\gamma_{k'}}\hat{S}_k(\Delta t)\hat{S}_{k'}(\Delta t)\hat{\rho}_S(0)\ket{0}\bra{0}\Bigr)\Bigr] = \mathbb{E}_{C}\bigl[\Tr_{E}\Bigl(\sqrt{\gamma_k}\sqrt{\gamma_{k'}}\hat{\rho}_S(0)\ket{0}\bra{0}\hat{S}_k(\Delta t)\hat{S}_{k'}(\Delta t)\Bigr)\Bigr] = 0.
\end{aligned}
\end{equation}
since the Wiener processes associated to different $k$ are independent.

\section{Approximation error due to the perturbative expansion}
\label{approx_error_appendix}
We derive an estimate of the upper bound on the approximation error due to the perturbative expansion used in the main text. To estimate the bound we adopt the superoperator norm $\norm{\mathcal{E}}_{1\rightarrow 1}\equiv \sup_{\norm{\hat{O}}_{1}=1}\norm{\mathcal{E}(\hat{O})}_{1}$ \cite{kliesch2011dissipative,sweke2015universal} where $\norm{\hat{O}}_{1} \equiv \Tr(\sqrt{\hat{O}^{\dagger}\hat{O}})$ and $\mathcal{E}(\, \cdot \,)$ is a generic superoperator. Moreover we use the fact that a formal solution of the Lindblad equation in interaction picture can be written as $\hat{\rho}_{S}(t+\Delta t) = \hat{U}(\Delta t)\text{T}\Bigl[e^{\int_t^{t+\Delta t}\D s\mathcal{D}(s)}\Bigr]\hat{\rho}_{S}(t)\hat{U}^{\dagger}(\Delta t)$, where  $\text{T}[\,\cdot\,]$ is the time ordering operator.

The approximation error is quantified by assuming a $m$-locality condition, namely by considering the decomposition of $\mathcal{D}(s)$ into $\mathcal{D}(s) = \sum_{j}^{K}\mathcal{D}_j(s)$ where each $\mathcal{D}_j(s)$ acts non trivially on a subset of $m < n$ qubits. Each of $\mathcal{D}_j(s)$ has a maximum of $2^{2m}-1$ Lindblad operators. For simplicity we assume that all the parameters $\gamma_k$ have the same order of magnitude $\gamma_k = \gamma$. We define the unitary superoperator $\mathcal{U}(\Delta t)(\hat{\rho}_S) = \hat{U}(\Delta t)\hat{\rho}_S\hat{U}^{\dagger}(\Delta t)$. Given the the approximate expression of the density matrix we found in the main text the upper bound on the approximation error can be computed as
\begin{equation}
\label{mostro1}
\begin{aligned}
\varepsilon_p &= \norm{\hat{U}(\Delta t)\text{T}\Bigl[e^{\int_t^{t+\Delta t}\D s\mathcal{D}(s)}\Bigr]\hat{U}^{\dagger}(\Delta t) - \hat{U}(\Delta t)\Bigl(\mathbb{1} +\int_t^{t+\Delta t}\D s\mathcal{D}(s)\Bigr)\hat{U}^{\dagger}(\Delta t)}_{1\rightarrow 1} \\
&\leq \norm{\mathcal{U}(\Delta t)}_{1\rightarrow 1}\cdot\norm{\text{T}\Bigl[e^{\int_t^{t+\Delta t}\D s\mathcal{D}(s)}\Bigr] - \Bigl(\mathbb{1} +\int_t^{t+\Delta t}\D s\mathcal{D}(s)\Bigr)}_{1\rightarrow 1}\\
&= \norm{\text{T}\Bigl[e^{\int_t^{t+\Delta t}\D s\mathcal{D}(s)}\Bigr] - \Bigl(\mathbb{1} +\int_t^{t+\Delta t}\D s\mathcal{D}(s)\Bigr)}_{1\rightarrow 1} \\
&= \norm{\sum_{k = 2}^{\infty}\frac{1}{k!}\int_{t}^{t+\Delta t}\dots \int_{t}^{t+\Delta t}\D t_1\dots \D t_k \text{T}\Bigl[\mathcal{D}(t_1)\dots \mathcal{D}(t_k)\Bigr]}_{1\rightarrow 1} \\
& \leq\sum_{k = 2}^{\infty}\frac{1}{k!}\norm{\int_{t}^{t+\Delta t}\dots \int_{t}^{t+\Delta t}\D t_1\dots \D t_k \text{T}\Bigl[\mathcal{D}(t_1)\dots \mathcal{D}(t_k)\Bigr]}_{1\rightarrow 1} = \sum_{k = 2}^{\infty}\frac{1}{k!}\norm{\text{T}\Biggl[\Bigl(\int_{t}^{t+\Delta t}\D s\mathcal{D}(s)\Bigr)^k\Biggr]}_{1\rightarrow 1}\\
&\leq\sum_{k = 2}^{\infty}\frac{1}{k!}\Bigl(\int_{t}^{t+\Delta t}\D s\norm{\mathcal{D}(s)}_{1\rightarrow 1}\Bigr)^k \leq  \frac{1}{2}\Bigl(\int_{t}^{t+\Delta t}\D s\norm{\mathcal{D}(s)}_{1 \rightarrow 1}\Bigr)^2\sum_{k = 0}^{\infty}\frac{1}{k!}\Bigl(\int_{t}^{t+\Delta t}\D s\norm{\mathcal{D}(s)}_{1\rightarrow 1}\Bigr)^k\\
&=  \frac{1}{2}\Bigl(\int_{t}^{t+\Delta t}\D s\norm{\mathcal{D}(s)}_{1 \rightarrow 1}\Bigr)^2e^{\int_{t}^{t+\Delta t}\D s\norm{\mathcal{D}(s)}_{1\rightarrow 1}}\leq \frac{e}{2}\Bigl(\int_{t}^{t+\Delta t}\D s\norm{\mathcal{D}(s)}_{1 \rightarrow 1}\Bigr)^2,
\end{aligned}
\end{equation}
where we use the fact that $\gamma \Delta t \ll 1$, thus $\int_{t}^{t+\Delta t}\D s\norm{\mathcal{D}(s)}_{1\rightarrow 1}< 1$ , the inequalities $\norm{\mathcal{E}_1+\mathcal{E}_2}_{1 \rightarrow 1} \leq \norm{\mathcal{E}_1}_{1 \rightarrow 1}+\norm{\mathcal{E}_2}_{1 \rightarrow 1}$, $\norm{\mathcal{E}_1\mathcal{E}_2}_{1 \rightarrow 1} \leq \norm{\mathcal{E}_1}_{1 \rightarrow 1}\norm{\mathcal{E}_2}_{1 \rightarrow 1}$ and $\norm{\mathcal{U}(\Delta t)}_{1\rightarrow 1} = 1$.

Now by considering the $m$-locality condition on $\mathcal{D}(s)$ described above, one can write $\norm{\mathcal{D}(s)}_{1\rightarrow 1}\leq K \max_{j} \norm{\mathcal{D}_j(s)}_{1\rightarrow 1}$. Thus, $\norm{\mathcal{D}_j(s)}_{1\rightarrow 1}$ can be expressed as
\begin{equation}
\label{mostro2}
\begin{aligned}
\norm{\mathcal{D}_j(s)}_{1\rightarrow 1} =& \sup_{\norm{\hat{O}}_1= 1}\norm{\sum_{k=1}^{2^{2m}-1}\gamma\Bigl( (\hat{L}_{k,j}(s)\hat{O}\hat{L}_{k,j}^{\dagger}(s)-\frac{1}{2}\{\hat{L}_{k,j}^{\dagger}(s)\hat{L}_{k,j}(s),\hat{O}\}\Bigr)}_{1}\\
&\leq \gamma\sum_{k=1}^{2^{2m}-1}\Bigl( \sup_{\norm{\hat{O}}_1= 1}\norm{\hat{L}_{k,j}(s)\hat{O}\hat{L}_{k,j}^{\dagger}(s)}_{1}+\frac{1}{2} \sup_{\norm{\hat{O}}_1= 1}\norm{\{\hat{L}_{k,j}^{\dagger}(s)\hat{L}_{k,j}(s),\hat{O}\}}_{1}\Bigr)\\
& \leq2\gamma\sum_{k=1}^{2^{2m}-1}\sup_{\norm{\hat{O}}_1= 1}\norm{\hat{O}}_{1}\norm{\hat{L}_{k,j}^{\dagger}(s)}_{\infty}\norm{\hat{L}_{k,j}(s)}_{\infty}\leq 2\gamma\sum_{k=1}^{2^{2m}-1}\norm{\hat{L}_{k,j}^{\dagger}(s)}_{\infty}\norm{\hat{L}_{k,j}(s)}_{\infty}\\
& =  2\gamma\sum_{k =1}^{2^{2m}-1}\norm{\hat{L}_{k,j}(s)}^{2}_{\infty} \leq 2 \gamma(2^{2m}-1)\max_k\Bigl(\norm{\hat{L}_{k,j}(s)}^{2}_{\infty} \Bigr)\\
& = 2\gamma (2^{2m}-1)\max_k\Bigl(\norm{\hat{L}_{k,j}}^{2}_{\infty} \Bigr),
\end{aligned}
\end{equation}
where we use the inequalities $\norm{\hat{A}\hat{B}}_{1} \leq \norm{\hat{A}}_{1}\norm{\hat{B}}_{\infty}$, $\norm{\hat{A}\hat{B}}_{1} \leq \norm{\hat{A}}_{\infty}\norm{\hat{B}}_{1}$ and $\norm{\hat{A}}_{\infty} = \sup_{\norm{\ket{\psi}}=1}\norm{\hat{A}\ket{\psi}} = \sup_{\norm{\ket{\psi}}=1}\sqrt{\bra{\psi}\hat{A}^{\dagger}\hat{A}\ket{\psi}}$. Moreover we exploit the fact that $\norm{\hat{L}_{k,j}(s)}_{\infty}^{2} = \sup_{\norm{\ket{\psi}}=1}\bra{\psi}\hat{U}^{\dagger}(s,t)\hat{L}_{k,j}^{\dagger}\hat{L}_{k,j}\hat{U}(s,t)\ket{\psi} = \sup_{\norm{\ket{\psi'}}=1}\bra{\psi'}\hat{L}_{k,j}^{\dagger}\hat{L}_{k,j}\ket{\psi'} = \norm{\hat{L}_{k,j}}_{\infty}^{2}$. Finally by inserting Eq. \eqref{mostro2} in Eq. \eqref{mostro1} one finds
\begin{equation}
   \varepsilon_p \leq 2 e \Bigl(K (2^{2m}-1)\max_{k,j}\norm{\hat{L}_{k,j}}^{2}_{\infty}\gamma\Delta t\Bigr)^2.
\end{equation}
$K$ that in general takes the value $n!/m!(n-m)!$ scales polinomially as the system size $n$ goes to infinity and read
\begin{equation}
K \sim \frac{n^m}{m!e^m}\sim \mathcal{O}(n^m),
\end{equation}
where we use the Stirling formula.

\section{On the locality of the product terms in Eq. \eqref{Q_noisy_gate}}
\label{locality}
We notice that the Lindblad operators in general are strings of operators with the structure $\hat{L}_k = \hat{l}_k^{(\alpha)}... \hat{l}_k^{(\beta)}$ where $\alpha$, $\beta$ are integer numbers, $\hat{l}_k^{(q)}$ with $q\in [1,...,\alpha,...,\beta,...,n]$ are generic single-qubit operators applied to the q-th qubit.  In the case of $m$-locality of the Lindblad equation, the number of $\hat{l}_k^{(q)}$ in $\hat{L}_k$ is $m<n$. Thus, by using the first order trotter-Suzuki formula $\hat{U}(\Delta t)\simeq\prod_\alpha e^{-\frac{i}{\hbar}\hat{H}_\alpha \Delta t}$, each operator $\hat{L}_k(s) = \hat{U}^{\dagger}(s-t)\hat{L}_k\hat{U}(s-t)$ in Eq. \eqref{Q_noisy_gate} has a fixed locality. Indeed $\hat{L}_k$ acts on a set $\mathbb{A}$ of $m$ qubits, then all the terms in $\hat{U}$ that act on a set of qubits disjoint from $\mathbb{A}$, do not contribute in the expression for $\hat{L}_k(s)$. As a result, the locality of $\hat{L}_k(s)$ depends only on $m$ and not on $n$ (in general if the locality of $\hat{H}_\alpha$ is $l\neq m$ then the locality of $\hat{L}_k(s)$ depends on $m$ and $l$). A remarkable property of the approach is that we can implement this by using the same unique ancillary bath qubit that has to be reset only after the application of all operators in the right hand side of Eq. (\ref{trick}). 

\section{Total approximation error of a single time step}\label{total_error_deltat}
Given the implementation of the algorithm in Sec. \ref{sec:algorithm} in this appendix we compute the total approximation error $\varepsilon$ for a single time step $\Delta t$. Here we define $\hat{U}_1(s-t) = \prod_{\alpha=1}^K e^{-\frac{i}{\hbar}\hat{H}_\alpha (s-t)}$ the first order Trotter-Suzuki product formula \cite{suzuki1985decomposition} and $\mathcal{D}_1(s) = \hat{U}_1^{\dagger}(s-t)\mathcal{D}\hat{U}_1(s-t)$. In the following we use the Zassenhaus formula $\hat{U}(s-t) = \hat{U}_1(s-t) e^{-\frac{1}{2}\sum_{\alpha<\beta}^K[\hat{H}_\alpha,\hat{H}_\beta](s-t)^2} + o((s-t)^3)$ \cite{suzuki1977convergence}.
\begin{equation}
\label{mostro3}
\begin{aligned}
\varepsilon &= \norm{\hat{U}(\Delta t)\text{T}\Bigl[e^{\int_t^{t+\Delta t}\D s\mathcal{D}(s)}\Bigr]\hat{U}^{\dagger}(\Delta t) - \hat{U}_1(\Delta t)\Bigl(\mathbb{1} +\int_t^{t+\Delta t}\D s\mathcal{D}_1(s)\Bigr)\hat{U}_1^{\dagger}(\Delta t)}_{1\rightarrow 1} \\
&=\Biggl\lVert\hat{U}(\Delta t)\text{T}\Bigl[e^{\int_t^{t+\Delta t}\D s\mathcal{D}(s)}\Bigr]\hat{U}^{\dagger}(\Delta t) - \hat{U}(\Delta t)\Bigl(\mathbb{1} +\int_t^{t+\Delta t}\D s\mathcal{D}(s)\Bigr)\hat{U}^{\dagger}(\Delta t)+ \hat{U}(\Delta t)\Bigl(\mathbb{1} +\int_t^{t+\Delta t}\D s\mathcal{D}(s)\Bigr)\hat{U}^{\dagger}(\Delta t)\\
&\qquad- \hat{U}_1(\Delta t)\Bigl(\mathbb{1} +\int_t^{t+\Delta t}\D s\mathcal{D}_1(s)\Bigr)\hat{U}_1^{\dagger}(\Delta t)\Biggr\rVert_{1\rightarrow 1}\\
&\leq \norm{\hat{U}(\Delta t)\text{T}\Bigl[e^{\int_t^{t+\Delta t}\D s\mathcal{D}(s)}\Bigr]\hat{U}^{\dagger}(\Delta t) - \hat{U}(\Delta t)\Bigl(\mathbb{1} +\int_t^{t+\Delta t}\D s\mathcal{D}(s)\Bigr)\hat{U}^{\dagger}(\Delta t)}_{1\rightarrow 1}\\
&\qquad+ \norm{\hat{U}(\Delta t)\Bigl(\mathbb{1} +\int_t^{t+\Delta t}\D s\mathcal{D}(s)\Bigr)\hat{U}^{\dagger}(\Delta t)-\hat{U}_1(\Delta t)\Bigl(\mathbb{1} +\int_t^{t+\Delta t}\D s\mathcal{D}_1(s)\Bigr)\hat{U}_1^{\dagger}(\Delta t)}_{1\rightarrow 1}\\
&\leq \varepsilon_p + \norm{\hat{U}(\Delta t)\Bigl(\mathbb{1} +\int_t^{t+\Delta t}\D s\mathcal{D}(s)\Bigr)\hat{U}^{\dagger}(\Delta t)-\hat{U}_1(\Delta t)\Bigl(\mathbb{1} +\int_t^{t+\Delta t}\D s\mathcal{D}_1(s)\Bigr)\hat{U}_1^{\dagger}(\Delta t)}_{1\rightarrow 1}.
\end{aligned}
\end{equation}
At this point we exploit the Zassenhaus formula truncated at second order $\hat{U}(s-t) \simeq \hat{U}_1(s-t)\hat{R}(s-t)$, where $\hat{R}(s-t) = e^{-\frac{1}{2}\sum_{\alpha<\beta}^K[\hat{H}_\alpha,\hat{H}_\beta](s-t)^2} \simeq \mathbb{1} -\frac{1}{2}\sum_{\alpha<\beta}^K[\hat{H}_\alpha,\hat{H}_\beta](s-t)^2= \mathbb{1} -\frac{\omega^2}{2}\sum_{\alpha<\beta}^K\sum_{j,j'=1}^{J}[\hat{h}_{\alpha,j},\hat{h}_{\beta,j'}](s-t)^2$. In the latter we defined $\hat{H}_\alpha = \omega\sum_{j=1}^J, \hat{h}_{\alpha,j}$ where $J$ is a constant whose value depends on the system Hamiltonian under study, $\hat{h}_{\alpha,j}$ are generic $m$-local operators and for simplicity we choose the same frequency $\omega$ for each $\hat{H}_\alpha$.  By defining the superoperator $\mathcal{U}_1(\Delta t)(\hat{\rho}_S) = \hat{U}_1(\Delta t)\hat{\rho}_S\hat{U}_1^{\dagger}(\Delta t)$ and by plugging the above expressions in Eq. \eqref{mostro3} we can further bound $\varepsilon$ as

\begin{equation}
\begin{aligned}
\varepsilon &\leq \varepsilon_p + \norm{\mathcal{U}_1(\Delta t)}_{1\rightarrow 1}\cdot\norm{\hat{R}(\Delta t)\Bigl(\mathbb{1} +\int_t^{t+\Delta t}\D s\hat{R}^{\dagger}(s-t)\mathcal{D}_1(s)\hat{R}(s-t)\Bigr)\hat{R}^{\dagger}(\Delta t)-\Bigl(\mathbb{1} +\int_t^{t+\Delta t}\D s\mathcal{D}_1(s)\Bigr)}_{1\rightarrow 1}\\
&\leq\varepsilon_p + \frac{(\omega\Delta t)^2}{2}\sum_{\alpha<\beta}^K\sum_{j,j'=1}^{J}\Biggl(\sup_{\norm{\hat{O}}=1}\norm{[\hat{h}_{\alpha,j},\hat{h}_{\beta,j'}]\hat{O}}_{1}+\sup_{\norm{\hat{O}}=1}\norm{\hat{O}[\hat{h}_{\alpha,j},\hat{h}_{\beta,j'}]}_{1}\Biggr) + o(\gamma\omega^2 (\Delta t)^3)\\
&\leq \varepsilon_p + (\omega\Delta t)^2\sum_{\alpha<\beta}^K\sum_{j,j'=1}^{J}\norm{[\hat{h}_{\alpha,j},\hat{h}_{\beta,j'}]}_{\infty}+ o(\gamma\omega^2 (\Delta t)^3)\\
&\leq \varepsilon_p + K^2J^2\max_{\alpha,\beta,j,j'}\norm{[\hat{h}_{\alpha,j},\hat{h}_{\beta,j'}]}_{\infty}(\omega\Delta t)^2+o(\gamma\omega^2 (\Delta t)^3)\\
&\leq \varepsilon_p + \Bigl(KJ\max_{\alpha,j}\norm{\hat{h}_{\alpha,j}}_{\infty} \omega\Delta t\Bigr)^2+o(\gamma\omega^2 (\Delta t)^3)=\varepsilon_p + \varepsilon_T +o(\gamma\omega^2 (\Delta t)^3)
\end{aligned}
\end{equation}

\section{Analysis of the approximation error for higher order Trotter-Suzuki product formula}\label{higher_trotter}

In Section \ref{sec:algorithm}
we presented the algorithm by using a first order Trotter-Suzuki product formula, however one can improve the approximation error to higher orders. By following a similar derivation as that of Appendix \ref{total_error_deltat} one obtains

\begin{equation}
\label{e_global_higher}
\varepsilon_{global}\leq N_{step} \Biggl(\varepsilon_p + \Bigl(KJ\max_{\alpha,j}\norm{\hat{h}_{\alpha,j}}_{\infty} \omega\Delta t\Bigr)^{\kappa+1}+o(\gamma\omega^{\kappa+1} (\Delta t)^{k+2})\Biggr),
\end{equation}
where $\kappa$ is the order of the Trotter-Suzuki product formula. We notice that in order to get the upper bound in Eq. \eqref{e_global_higher} one has to compute
\begin{equation}
\varepsilon = \norm{\hat{U}(\Delta t)\text{T}\Bigl[e^{\int_t^{t+\Delta t}\D s\mathcal{D}(s)}\Bigr]\hat{U}^{\dagger}(\Delta t) - \hat{U}_\kappa(\Delta t)\Bigl(\mathbb{1} +\int_t^{t+\Delta t}\D s\mathcal{D}_\kappa(s)\Bigr)\hat{U}_\kappa^{\dagger}(\Delta t)}_{1\rightarrow 1},
\end{equation}
where we define $\hat{U}_\kappa(s-t)$ the $\kappa$-th order Trotter-Suzuki product formula and $\mathcal{D}_\kappa(s) = \hat{U}_\kappa^{\dagger}(s-t)\mathcal{D}\hat{U}_\kappa(s-t)$. However one gets the same upper bound by computing
\begin{equation}
\varepsilon = \norm{\hat{U}(\Delta t)\text{T}\Bigl[e^{\int_t^{t+\Delta t}\D s\mathcal{D}(s)}\Bigr]\hat{U}^{\dagger}(\Delta t) - \hat{U}_\kappa(\Delta t)\Bigl(\mathbb{1} +\int_t^{t+\Delta t}\D s\mathcal{D}_1(s)\Bigr)\hat{U}_\kappa^{\dagger}(\Delta t)}_{1\rightarrow 1}.
\end{equation}
Thus in Alg. \ref{alNG} the $\hat{U}(s-t)$ used to implement the stochastic terms $\hat{S}_k(\Delta t)$ in Eq. \eqref{Q_noisy_gate} can be chosen as $\hat{U}_1(s-t)$ instead of $\hat{U}_\kappa(s-t)$, which is computationally more efficient.

\section{Domain of application}
\label{domain_application_appendix}
The advantage of our protocol is that it is based on a perturbative expansion of the environment coupling constants for which $\Delta t$ is not restricted to be infinitesimal. We provide an approximate solution of the Lindblad equation (Eq. \eqref{rho_ng_approx}) which is more accurate in the interval $\Delta t$ than just 
\begin{equation}
\label{rho_infinitesimal}
\hat{\rho}_{S}(t+\Delta t) =\hat{\rho}_{S}(t)+\Bigl(-\frac{i}{\hbar}[\hat{H}_{S},\hat{\rho}_{S}(t)]+\mathcal{D}\hat{\rho}_{S}(t)\Bigr)\Delta t
\end{equation}
that instead requires an infinitesimal $\Delta t$ to be valid. Thus, given a fixed target accuracy $\varepsilon_{global}$, when the perturbative expansion holds ($\gamma << \omega$) our approximation reduces the number of total time steps required to reach the final time $T$ with respect to Eq. \eqref{rho_infinitesimal}. This is also confirmed by the panel (a) in Fig. \ref{trace_distance}.

Moreover, our upper bound on $\varepsilon_{global}$ (Eq. \eqref{e_global}) and on the number of gates $\text{N}_G$ (Eq. \eqref{stima_numero_gates}) are smaller than those of other approaches \cite{cattaneo2021collision, cattaneo2023quantum,pocrnic2024quantum}. In particular our upper bounds contain less terms to leading order in $\gamma \Delta t$ and $\omega \Delta t$.

We mention that when the characteristic times of the environmental noise and of the system Hamiltonian evolution become comparable, i.e. $\gamma \sim \omega$, the perturbative approach is not justified anymore. In this regime, Eq. \eqref{rho_ng_approx} reduces to Eq. \eqref{rho_infinitesimal} and the stochastic exponents in $\hat{N}(\Delta t)  \simeq  \hat{U}(\Delta t) \prod_k e^{ \sqrt{\gamma_k}\hat{S}_k(\Delta t)}$ become $\sqrt{\gamma}_k\hat{J}_k W_k(\Delta t)$, where $\hat{J}_k \equiv \hat{L}_k \hat{\sigma}_{E}^{+} - \hat{L}_k^{\dagger} \hat{\sigma}_{E}^{-}$. As $W_k(\Delta t)$ have variances $\Delta t$, our algorithm resembles a stochastic implementation of the algorithm proposed in \cite{cattaneo2021collision,cattaneo2023quantum} with the significant difference that in our case the system-environment interaction is mediated by a single ancillary qubit regardless of the total number of the system Lindblad operators.

\section{Sampling error}
\label{sampling_error_appendix}
As it is known in the literature \cite{Molmer:93,daley2014quantum,bonnes2014superoperators}, the estimate of the expectation value of an observable $\hat{O}$ with finite number of realization of the stochastic processes have a sampling error of the form 
\begin{equation}\label{sampling_error_eta}
    \eta = |\langle\hat{O}\rangle-\langle\hat{O}\rangle_{\mathcal{N}_{r}}| = \frac{\Delta \hat{O}(\gamma, \Delta t)}{\sqrt{\mathcal{N}_r}}
\end{equation}
where $\langle\hat{O}\rangle = \Tr (\hat{O}\hat{\rho}_S(t+\Delta t))$ is the expected value in the limit of infinite samples,  where $\hat{\rho}_{S}(t+\Delta t) =\mathbb{E}_{C}\bigl[\Tr_{E}(\hat{N}(\Delta t)\ket{\Psi(t)}\bra{\Psi(t)}\hat{N}^{\dagger}(\Delta t))\bigr]$ and $\langle\hat{O}\rangle_{\mathcal{N}_{r}} = \Tr (\hat{O}\hat{\rho}_{\mathcal{N}_{r}}(t+\Delta t))$ is the estimate with $\mathcal{N}_{r}$ total number of samples of the classical stochastic processes where $\hat{\rho}_{\mathcal{N}_{r}}(t+\Delta t) = \frac{1}{\mathcal{N}_{r}}\sum_{k}^{\mathcal{N}_{r}}\Tr_{E}(\hat{N}_k(\Delta t)\ket{\Psi(t)}\bra{\Psi(t)}\hat{N}_k^{\dagger}(\Delta t))$. The value $\Delta \hat{O}(\gamma, \Delta t)$ is the square root of the sample variance $\Delta \hat{O}^2(\gamma, \Delta t) = \frac{1}{\mathcal{N}_{r}}\sum_{k}^{\mathcal{N}_{r}}\Tr^2 ( \hat{O} \Tr_{E}(\hat{N}_k(\Delta t)\ket{\Psi(t)}\bra{\Psi(t)}\hat{N}_k^{\dagger}(\Delta t)))  -\langle\hat{O}\rangle^2_{\mathcal{N}_{r}}$ which is upper bounded by (see \cite{Molmer:93})
\begin{equation}\label{bound_variance_error}
   \Delta \hat{O}^2(\gamma, \Delta t) \leq \langle\hat{O}^2\rangle_{\mathcal{N}_{r}} -\langle\hat{O}\rangle^2_{\mathcal{N}_{r}} \simeq \langle\hat{O}^2\rangle -\langle\hat{O}\rangle^2.
\end{equation}

It can be shown \cite{Molmer:93} that for global operators $\hat{O}$ the following relation holds
\begin{equation}
    \sqrt{\langle\hat{O}^2\rangle -\langle\hat{O}\rangle^2} \sim \langle\hat{O}\rangle,
\end{equation}
thus by using Eqs. \eqref{sampling_error_eta} and \eqref{bound_variance_error} we get 
\begin{equation}
    \langle\hat{O}\rangle \pm \eta \leq \langle\hat{O}\rangle\biggl(1\pm 1/\sqrt{\mathcal{N}_{r}}\biggr),
\end{equation}
showing that the numbers of samples needed to reach a target accuracy does not depend on the dimension of the system.

\section{Values of the coupling constants $\gamma_k$ used in Fig. \ref{fig:two_qubits}}
\label{coupling_constants_values}
Here we report a table with the numeric values of the coupling constants used to reproduce the evolution in Fig. \ref{fig:two_qubits}.

\begin{table}[h]
\centering
\begin{tabular}{|c|c|c|c|c|c|c|c|c|c|c|c|c|c|c|}
\hline
$\hat{I}\hat{X}$  & $\hat{I}\hat{Y}$ & $\hat{I}\hat{Z}$
& $\hat{X}\hat{I}$ & $\hat{Y}\hat{I}$ & $\hat{Z}\hat{I}$ & $\hat{X}\hat{X}$& $\hat{X}\hat{Y}$&$\hat{X}\hat{Z}$&$\hat{Y}\hat{X}$&$\hat{Y}\hat{Y}$&$\hat{Y}\hat{Z}$&$\hat{Z}\hat{X}$&$\hat{Z}\hat{Y}$&$\hat{Z}\hat{Z}$\\
\hline
0.005    & 0.034 & 0.300        
& 0.250 & 0.096 & 0.280& 0.044&0.099&0.040&0.030&0.060&0.084&0.000&0.000&0.099\\
\hline
\end{tabular}
\caption{Values of the coupling constants for each Pauli string.}
\label{tab:my_label}
\end{table}

\bibliography{biblio_resub}

\end{document}